\documentclass[11pt,letterpaper]{article}

\usepackage{mathpazo}

\usepackage{authblk}

\usepackage{amsmath, amsthm, amssymb,color,fullpage,url,booktabs}  
\usepackage[pdftex]{hyperref} 
\usepackage{graphicx}

\newcommand{\nc}{\newcommand}
\nc{\rnc}{\renewcommand}

\newcommand{\bra}[1]{\left\langle #1\right|}
\newcommand{\ket}[1]{\left|#1\right\rangle}

\def\be#1\ee{\begin{equation}#1\end{equation}}
\def\bea#1\eea{\begin{eqnarray}#1\end{eqnarray}}
\def\beas#1\eeas{\begin{eqnarray*}#1\end{eqnarray*}}
\def\ba#1\ea{\begin{align}#1\end{align}}
\def\bas#1\eas{\begin{align*}#1\end{align*}}
\def\bpm#1\epm{\begin{pmatrix}#1\end{pmatrix}}

\newtheorem{thm}{Theorem}
\newtheorem*{thm*}{Theorem}

\newtheorem{dfn}[thm]{Definition}
\newtheorem{proto}{Protocol}

\makeatletter
\newtheorem*{rep@theorem}{\rep@title}
\newcommand{\newreptheorem}[2]{%
\newenvironment{rep#1}[1]{%
 \def\rep@title{#2 \ref{##1} (restatement)}%
 \begin{rep@theorem}}%
 {\end{rep@theorem}}}
\makeatother

\newreptheorem{thm}{Theorem}
\newreptheorem{lem}{Lemma}

\def\benum{\begin{enumerate}}
\def\eenum{\end{enumerate}}
\def\bit{\begin{itemize}}
\def\eit{\end{itemize}}

\nc{\todo}[1]{\textcolor{red}{todo: #1}}

\def\begsub#1#2\endsub{\begin{subequations}\label{eq:#1}\begin{align}#2\end{align}\end{subequations}}
\nc\qand{\qquad\text{and}\qquad}
\nc\mnb[1]{\medskip\noindent{\bf #1}}

\usepackage{boxedminipage}

\usepackage{amsmath}
\usepackage{amssymb}

\usepackage[T1]{fontenc}
\usepackage[utf8]{inputenc}
\usepackage{authblk}

\usepackage{enumerate}
\usepackage{enumitem}
\usepackage[]{units}
\usepackage{amsfonts}
\usepackage{amsmath}
\usepackage{mathtools}
\usepackage{amsthm}
\usepackage{bm}
\usepackage{mciteplus}
\usepackage[numbers]{natbib}

\newtheorem*{lem1}{Lemma 1}
\newtheorem*{lem2}{Lemma 2}
\newtheorem*{lem3}{Lemma 3}
\newtheorem*{lem4}{Lemma 4}
\newtheorem*{lem5}{Lemma 5}
\newtheorem*{lem6}{Lemma 6}
\newtheorem*{lem7}{Lemma 7}

\newtheorem{theorem}{Theorem}

\newtheorem{corollary}[theorem]{Corollary}

\newtheorem{definition}[theorem]{Definition}

\newcommand{\ii}{\mathbb{I}}											

\newcommand{\tur}{\,\circlearrowleft\,}   
\newcommand{\gat}{\,\blacktriangleleft}  
\newcommand{\mov}{\,\vartriangleleft}    
  
\newcommand{\rmov}{\,\vartriangleright}
   
\newcommand{\bul}{\:\:\centerdot\:}       
\newcommand{\iga}{\:I\,}                  

\newcommand{\li}{\overleftarrow{I}}
\newcommand{\ri}{\overrightarrow{I}}

\newcommand{\wga}{W}						

\newcommand{\lw}{\overleftarrow{W}}
\newcommand{\rw}{\overrightarrow{W}}

\newcommand{\sga}{\:S\,}					

\newcommand{\ls}{\overleftarrow{S}}
\newcommand{\rs}{\overrightarrow{S}}
\newcommand{\aga}{\,A\,}					

\newcommand{\la}{\overleftarrow{A}}
\newcommand{\bga}{\,B\,}					

\newcommand{\lb}{\overleftarrow{B}}
\newcommand{\rb}{\overrightarrow{B}}
\newcommand{\goes}{\:\rightarrow\:}		
\newcommand{\band}[2]{		
	\begin{array}{|r|r|}
	\hline #1 & #2 \\
	\hline
	\end{array}}

\newcommand{\triUL}[3]{ 		
	\:\:\begin{array}{|c|c@{}l}
	\cline{1-2} #1 & #2 \:\: & \vline \\ 
	\cline{1-2} #3 & & \\ 
	\cline{1-1}
	\end{array}}

\newcommand{\eight}[8]{ 		
	\begin{array}{|c|c|}
	\hline #1 & #2 \\
	\hline #3 & #4 \\
	\hline #5 & #6 \\
	\hline #7 & #8 \\
	\hline
	\end{array}}
\newcommand{\ten}[8]{ 		
	\begin{array}{|c|c|}
	\hline #1 & #2 \\
	\hline - & - \\
	\hline #3 & #4 \\
	\hline #5 & #6 \\
	\hline #7 & #8 \\
	\hline
	\end{array}}
\newcommand{\twelve}[6]{ 		
	\begin{array}{|c|c|}
	\hline #1 & #2 \\
	\hline - & - \\
	\hline - & - \\
	\hline #3 & #4 \\
	\hline - & - \\
	\hline #5 & #6 \\
	\hline
	\end{array}}

\newcommand{\flour}[3]{ 		
	\begin{array}{|c|c|}
	\hline #1 & #2 \\
	\hline \multicolumn{2}{|c|}{#3} \\
	\hline
	\end{array}}

\newlength{\onebox}
\newcommand\raiseonebox{\raisebox{-.5\onebox} 
  {\rule{0pt}{\onebox}}}

\begin{document}

\title{Universal Hamiltonians for Exponentially Long Simulation}

\author{Thomas C. Bohdanowicz}

\author{Fernando G.S.L. Brand\~ao}

\affil{Institute of Quantum Information and Matter, California Institute of Technology, Pasadena, CA}

\date{\today}

\maketitle

\begin{abstract}
We construct a Hamiltonian whose dynamics simulate the dynamics of every other Hamiltonian up to exponentially long times in the system size. The Hamiltonian is time independent, local, one dimensional, and translation invariant. As a consequence, we show (under plausible computational complexity assumptions) that the circuit complexity of the unitary dynamics under this Hamiltonian grows steadily with time up to an exponential value in system size. This result makes progress on a recent conjecture by Susskind, in the context of the AdS/CFT correspondence, that the time evolution of the thermofield double state of two conformal field theories with a holographic dual has circuit complexity increasing linearly in time, up to exponential time.
\end{abstract}

\section{Introduction}

The notion of \textit{universality} is central in science. At its core is the idea that a simple set of objects can describe the fundamental properties of a much richer class of objects. One particular successful example in physics is the use of universality in classifying phase transitions. It turns out that critical phenomena can be understood by only looking at a small set of systems and critical exponents associated to them, which are universal in the sense that they can reproduce the critical behavior of every other system \cite{IsingKad}. A second example, which is the focus of this paper, is the study of universal quantum dynamics. The goal here is to understand and classify, in several different scenarios, which quantum dynamics can simulate any other \cite{cubmont}. This line of investigation is at the core of the idea of a (universal) quantum computer, which should be able to efficiently simulate the dynamics of any quantum system \cite{feynman,deutsch}. It also naturally extends to the quantum domain the study of universality in classical dynamics, usually considered in the context of cellular automata. 

There are several different notions of universality of quantum dynamics. One of the first to be considered was the notion of universal quantum gates, in which one is interested in sets of quantum gates that can approximate any unitary operator. This is a rich problem and there is a beautiful theory around it \cite{KitSol,DN}, still with many unresolved questions. A noteworthy result is that almost any two qubit gate is universal in the sense that it (together with the ability to swap the qubits) can approximate any other unitary. Therefore, in a well defined sense, universality is the general rule for quantum gates. 

Another notion concerns universal Hamiltonians. Here we are interested in identifying Hamiltonians whose dynamics can simulate any other (more precisely we can ask whether we can perform universal quantum computation with the Hamiltonian). One variant of the question is to consider sets of few-qubit interactions and classify which are universal, given the ability of switching the interactions in a time-dependent fashion. This question is closely related to the one for quantum gates and, similarly, a generic two-qubit Hamiltonian is typically universal for quantum computing \cite{childs2010characterization}. 

Another interesting variant is to consider a fixed time-independent Hamiltonian and ask if it is universal. It is clear that the class of unitaries which can be approximated by the time evolution of a fixed Hamiltonian is rather limited (as the eigenbasis of $e^{i t H}$ for a fixed $H$ is always the same). However the situation is more interesting if one allows for some form of encoding of the dynamics by preparing the initial state of the system in a suitable way. Indeed, Vollbrecht and Cirac \cite{VC} gave a construction of a time-independent and translation-invariant model which is universal for quantum computing. The desired quantum circuit to be implemented is encoded in the initial state as a particular computational basis state. Later on, Nagaj and Wocjan made refinements to this construction in \cite{NW}. In contrast to the other notions of universality, it is unclear if a typical local Hamiltonian will be universal in the sense above. Indeed apart from a few examples, it is still largely unexplored which Hamiltonians are universal for time-independent evolutions. 

\vspace{0.2 cm}

\noindent \textbf{Circuit Complexity:} An a priori unrelated question in quantum complexity theory concerns the circuit complexity of quantum dynamics. The circuit complexity of a unitary operator consists of the minimum number of two-qubit gates needed to (approximately) construct the unitary. Large complexity is a general feature of unitary operators. Indeed a simple counting argument shows that most unitaries on $n$ qubits will have circuit complexity $2^{\Omega(n)}$ \cite{AB}. Moreover in Ref. \cite{brandao2012local} it was shown that most quantum circuits with $m$ gates have circuit complexity at least $\Omega(m^{1/11})$. It is much harder to prove circuit complexity lower bounds for the evolution of a fixed Hamiltonian (or even for the application of a fixed unitary many times). Indeed, proving a superpolynomial lower bound on circuit complexity for exponential long times would result in a major breakthrough in classical complexity theory (as shown in \cite{AB}, it would imply that $\textsc{PSPACE}$ is not contained in $\textsc{BQP}/$poly, which appears to be far beyond the reach of current proof techniques). 

It turns out that the concept of universality naturally connects with circuit complexity. If one aims to find a unitary whose dynamics has large circuit complexity, it would be natural to consider one whose time evolution is universal, capable of simulating any other evolution (with a reasonable overhead). Indeed, from the result of \cite{brandao2012local}, it directly follows that we can lower bound the circuit complexity of the unitary evolution of the Vollbrecht-Cirac universal Hamiltonian \cite{VC} on $n$ qubits, for large enough time, by $n^{c}$, for a constant $c > 0$. However it is not clear how to obtain larger lower bounds (i.e. superpolynomial in $n$) for long times, e.g. exponential in $n$, even under computational complexity assumptions. This is a natural problem to consider on its own (e.g. it was considered in Ref. \cite{atia2016fast} under the name of "fast-forwarding" and connected to the time-energy uncertainty relation), but it has recently gained a renewed interest due to a possible unexpected application in the context of quantum gravity and holography, which we now briefly review.  

\vspace{0,2 cm}

\noindent \textit{Susskind's Conjecture:} The AdS/CFT correspondence \cite{JM} posits the equivalence between a theory of quantum gravity in Anti-de-Sitter (AdS) space (i.e. gravity with a negative cosmological constant) with $d+1$ dimensions (called the bulk theory) and a quantum conformal field theory (CFT) with $d$ dimensions (called the boundary theory). This idea suggests that gravity is an emergent phenomenon (as it does not appear explicitly in the CFT description) and has become one of the most influential ideas in physics in the last twenty years. The correspondence is usually phrased as a dictionary between properties of the two theories. 

Yet there is one property of the quantum gravity picture for which it is hard to find a suitable partner in the CFT picture. This is the volume of a non-traversible wormhole (or Einstein-Rosen bridge) in AdS space connecting two boundary CFTs. One can argue that the volume of the wormhole will grow linearly with time up to exponential time \cite{LS}. But most  of the traditionally considered physical properties of the CFT (which usually involve only few-body operators) will equilibrate much sooner (on the timescale of the scrambling time). One property of the CFT system which does seem to have the same behaviour as the volume of the wormhole is the circuit complexity of the joint quantum state\footnote{Analogously to the circuit complexity of a unitary, the circuit complexity of a state is the minimum number of two-qubit gates (from a given universal set of gates) needed to (approximately) create the state.} of the two CFTs associated with the boundary of the wormhole (given by the time evolution of the so-called `thermofield double state' (TDS), which at infinite temperature reduces to a maximally entangled state of the two CFTs). Indeed, in analogy with the behavior of the wormhole volume, we expect the circuit complexity of the state to grow linearly with time up to exponential in the size of the system, at which point it has to saturate. Then at recursion times (which is doubly exponential in system size) the circuit complexity will have sharp oscillation before stabilizing again at an exponential value. Such a connection prompted Susskind to conjecture that the circuit complexity of the CFT state is the dual property of the volume of the wormhole.

Although fascinating, Susskind's proposal is still somewhat speculative; it is an interesting and challenging task to make it more concrete. Progress was made by Aaronson and Susskind \cite{SA}. They considered a toy model of the problem for which concrete results about the circuit complexity could be established. Let $V$ describe one step of the time evolution of the CFT. The CFT thermofield double state (at infinite temperature and regularizing the theory to have finite dimension $N$) after $t$ time steps is given by
\begin{equation}
\ket{\text{TDS}(t)} := \frac{1}{\sqrt{N}} \sum_i V^t \ket{i}_{\text{CFT, 1}} \otimes (V^T)^t \ket{i}_{\text{CFT, 2}} = \frac{1}{\sqrt{N}} \sum_i \ket{i}_{\text{CFT, 1}} \otimes U^t \ket{i}_{\text{CFT, 2}}
\end{equation} 
with $U = (V^T)^2$, where $V^T$ is the transpose of $V$. For Susskind's correspondence to hold, we need the circuit complexity of $\ket{\text{TDS}(t)}$ to grow linearly in time up to an exponential value in the system size. Aaronson and Susskind abstracted away the fact that $U$ is associated to the dynamics of a CFT  and asked if there is any (efficiently implementable) unitary $U$ for which one can show that the corresponding $\ket{\text{TDS}(t)}$ has complexity growing up to exponential with the number of applications $t$. They proved that choosing $U$ as a step function of a reversible and computationally-universal cellular automaton achieves the goal (under certain computational complexity assumptions). A natural open question is whether one can prove something similar for the evolution of the CFT. Short of that, can we get closer to this goal? For example, can we find a local Hamiltonian whose evolution can replace the step function of the cellular automaton in the Aaronson-Susskind reasoning?

\subsection{Main Results}

In this paper we consider universality for fixed Hamiltonians up to exponential times. All Hamiltonians will be normalized to have $\Vert H \Vert = 1$. Our desired notion of universality is captured by the following definition \footnote{This definition was suggested to us by Dorit Aharonov \cite{dorit}.}:

\begin{definition} \label{universal}
	A family of Hamiltonians $\{ H_m \}_{m \in \mathbb{N}}$, indexed by the number of qudits $m$ they act on, is called a \emph{Universal Hamiltonian Family} if for any $k$-local Hamiltonian $H\in\mathcal{B}(\mathbb{C}^{2^n})$ on $n$ qubits (with $k \leq O(\log(n))$) and time $t$, there are ${\rm{poly}}(n)$-sized quantum circuits $D$ and $E$, $m={\rm{poly}}(n,\log{t})$, and $t'={\rm{poly}}(t, n)$ such that
	\begin{eqnarray}
		\|e^{iHt}-  (I^{\otimes n} \otimes \bra{0^{m-n}})De^{iH_mt'}E(I^{\otimes n} \otimes \ket{0^{m-n}})  \|< 1/\rm{poly}(n). 
	\end{eqnarray}
\end{definition}

Previous results \cite{VC,NW} achieved a weaker notion of universality. In the notation of Definition \ref{universal}, the size $m$ of the Hamiltonians in Refs. \cite{VC,NW} scales as $m = {\rm{poly}}(n,t)$, which prevents efficient simulation for exponentially long times because it would require preparing an exponentially large state encoding the dynamics. So, in words, our construction allows exponentially long simulation time using only polynomial space and without requiring any active control or intervention during the simulation. Note also that this notion of universality is imcomparable to the notion considered in \cite{cubmont}. Rather than reproducing a large class of properties of the system being simulated (e.g. spectrum of Hamiltonian, etc.) our simulation scheme only simulates the \emph{dynamics} and is weaker in that sense. However, our simulation faithfully simulates dynamics for times up to exponentially large in the system size, whereas the construction of Ref. \cite{cubmont} can not, and is in that sense stronger.

Although Definition \ref{universal} is natural when discussing universal Hamiltonians, we can also consider the following stronger definition, which will be useful to us:
\begin{definition} \label{universalcircuit}
	A family of Hamiltonians $\{ H_m \}_{m \in \mathbb{N}}$, indexed by the number of qudits $m$ they act on, is called a \emph{Circuit Universal Hamiltonian Family} if for every ${\rm{poly}}(n)$-sized circuit $U$ on $n$ qubits and time $t$, there are ${\rm{poly}}(n)$-sized quantum circuits $D$ and $E$, $m={\rm{poly}}(n,\log{t})$, and $t'={\rm{poly}}(t, n)$ such that
	\begin{eqnarray}
		\| U^t-  (I^{\otimes n} \otimes \bra{0^{m-n}})De^{iH_mt'}E(I^{\otimes n} \otimes \ket{0^{m-n}})  \|< 1/\rm{poly}(n). 
	\end{eqnarray}
\end{definition}

From the Hamiltonian simulation results of \cite{berry2015hamiltonian}, it follows that a \emph{Circuit Universal Hamiltonian Family} is also a \emph{Universal Hamiltonian Family}. Our main result is the following:
\begin{theorem} \label{main}
There is a \emph{Circuit Universal Hamiltonian Family} in one spatial dimension with translation-invariance and local spin dimension of 14580. The encoding circuit consists of preparing computational basis product states. The decoding circuit consists of making a measurement in the computational basis and resetting computational basis product states to the all zero state. 
\end{theorem}

It is an open question to decrease the local dimension of the model; we expect substantial improvements to be possible. In words, we explicitly construct a local term $h_{l, l+1}$ acting on two qudits, each of dimension 14580, such that the family of Hamiltonians
\begin{equation}
H_{m} = \frac{1}{m} \sum_{l=1}^{m-1} h_{l, l+1}
\end{equation}
is universal in the sense of Definition \ref{universalcircuit}.

Theorem \ref{main} can be used to argue that the circuit complexity of $e^{iH_mt}$ must grow to a superpolynomial value. The starting point is the result of Atia and Aharonov that unless $\textsc{PSPACE} = \textsc{BQP}$, there must exist a 2-sparse row computable Hamiltonian whose long-time dynamics has superpolynomial circuit complexity  (see Theorem 6 of \cite{atia2016fast}). This result together with Theorem \ref{main} gives:

\begin{corollary} \label{growingcomplexity}
Unless $\textsc{PSPACE} = \textsc{BQP}$, the circuit complexity of $e^{i t H_m}$ is superpolynomial for $t = 2^{O(m)}$.   
\end{corollary}

An interesting feature of Corollary \ref{growingcomplexity} is that we can identify a \textit{single} Hamiltonian -- more precisely a fixed interaction term forming a translation-invariant model for each length -- which can be shown to have growing circuit complexity (under certain computational complexity assumptions). 

We can also make progress on Susskind proposal discussed before. Define 
\begin{equation}
\ket{\text{TDS}_{H_m}(t)} := \frac{1}{\sqrt{N}} \sum_i \ket{i}\otimes e^{i t H_m} \ket{i}.
\end{equation} 
Then we have:

\begin{corollary} \label{growingcomplexityTDS}
Unless $\textsc{PSPACE}$ is inside $\textsc{PP}/$poly, the circuit complexity of $\ket{\text{TDS}_{H_m}(t)}$ is superpolynomial for $t = 2^{O(m)}$. Assuming that there are no subexponential $\textsc{PP}$ circuits for $\textsc{PSPACE}$,  the circuit complexity of $e^{i t H_m}$ for $t = 2^{O(m)}$ is $2^{\Omega(m)}$.
\end{corollary}

The proof follows directly from Theorem 7.1.2 of \cite{AB}, by using $H_m$ to simulate the step function of the universal cellular automaton used there.  The corollary makes partial progress on Susskind's proposal, by finding a quantum evolution which shares more features with the dynamics of a CFT (locality and translation-invariance) than the cellular automaton of \cite{SA}. We leave as open questions demonstrations that our construction can replicate important features and symmetries expected of a CFT Hamiltonian, and whether or not it can be used to create a state whose circuit complexity actually grows \textit{linearly} in $t$ from $t=0$ up to exponential times (under plausible complexity theoretic assumptions).



\subsection{Sketch of Construction}

The construction of our circuit universal Hamiltonian family (CUHF) that proves Theorem \ref{main} makes use of the concept of a Hamiltonian Quantum Cellular Automaton (HQCA), as described by Nagaj and Wocjan in \cite{NW}. The formal definition of an HQCA is as follows:
\begin{definition}\label{HQCA}
	A Hamiltonian Quantum Cellular Automaton (HQCA) is a local, time-independent, translation-invariant Hamiltonian $H$ on a lattice of qudits which carries out quantum computation via the following sequence of steps:
	\begin{enumerate}
		\item The input of the computation as well as information describing the computation to be performed (which is described by some unitary operator $U$) is encoded in the state of the qudits.
		\item The qudits undergo continuous time evolution under $H$ for some time $t$.
		\item A simple basis state measurement on a subset of the qudits collapses, with high probability, the state of the whole system to one where an appropriate subset of the qudits contains the desired output of the quantum computation.
	\end{enumerate}
\end{definition}
Such a Hamiltonian $H$ is termed a cellular automaton because the local terms which sum up to make $H$ can usually be thought of as transitions between different basis states for a local part of the lattice, which themselves correspond to reversible transition rules of some classical cellular automaton. This will be evident in the description of our construction.

Our CUHF $\{H_m\}_{m\in\mathbb{N}}$ will actually be a family of universal HQCA on a 1$D$ chain of $m$ qudits, which satisfy the following properties in addition to those in the definition above:
\begin{itemize}
	\item The local qudit dimension is constant.
	\item A quantum circuit $U$ with $C$ gates acting on $N$ qubits can be applied $t$ times through time evolution via Hamiltonian $H_m$ on a chain of $m={\rm{poly}}(N,C,\log{t})$ qudits.
\end{itemize}
The starting point of our construction is that of Nagaj and Wocjan \cite{NW}, which produces a 1$D$ CUHF with constant local qudit dimension $d=20$, for which the size of the chain scales as $m={\rm{poly}}(N,C)$. A version of this construction is described in detail in section \ref{revisionNagajWocjan}, where we also identify and highlight some important features of their construction that are essential to preserve in our eventual final construction \footnote{E.g., the definition and consequences of having a uniquely orthogonally generated set of states for an HQCA, see section \ref{hamgeo}}.

We build an HQCA which only encodes $U$ into the input state and a number $t$ encoded in binary, but which is able to apply $U$ repeatedly up to $t$ times after waiting for a time $t'={\rm{poly}}(N, C, t)$ \cite{elizabeth}. Therefore the size of the system needed for the simulation is only $m={\rm{poly}}(N,\log{t})$. In the following sections, we proceed to augment Nagaj-Wocjan construction  \cite{NW} step by step to add in more functionality until we achieve the desired $m={\rm{poly}}(N,\log{t})$ scaling for the simulation size.

The innovations which allow us to obtain our final construction dictate the organization of the remainder of the paper:

\begin{enumerate}

\item The Nagaj-Wocjan construction can only apply its encoded unitary $U$ once. So, first, we modify the Nagaj-Wocjan construction so that when the HQCA reaches its end state where $U$ has been applied, it can undergo transitions to `reset' itself and allow for another application of $U$ without undoing the original application of $U$. This basically has the effect of allowing the Nagaj-Wocjan construction to run over and over again rather than just once. This is described in Section \ref{conii}. 

\item Next, we augment the construction to include a binary clock in the qudit chain which counts the number of times that $U$ has been applied. This allows states of the chain with different numbers of $U$ applications to be orthogonal so that they can be distinguished by a simple measurement. Binary clock constructions of a similar nature also appear in \cite{AGIK,GI,CPGW}, in the context of proving hardness/uncomputability results for the ground-energy/gap of translation-invariant models. This is described in Section \ref{coniii}.

\item Finally, we augment the construction with the addition of a target register that allows one to specify a certain desired number of applications of $U$ which is stored in the start state. Once the clock has counted this number of applications of $U$, the HQCA continues evolving forward in time similarly to the way in which it did before, but it no longer makes any applications of $U$ to the qubits. This allows one to have complete control over how many times $U$ is applied to the qubits, and is crucial to being able to guarantee that after waiting a reasonable amount of time, performing a simple measurement on a subset of the chain is likely to yield a state with the desired number of applications of $U$. This is described in Section \ref{coniv}.
\end{enumerate}

The main challenge in achieving each of these results was to carefully engineer transition rules such that starting from the appropriate start state, for every state that can be obtained from the start state through application of transition rules, exactly one forward transition rule and one reverse transition rule is applicable to that state. This guarantees that the sequence of states obtained from the start state through application of the HQCA transition rules can map directly onto position eigenstates of a single particle on a chain, so that the analysis of the time evolution under the HQCA reduces to that of a single particle continuous-time quantum walk on a line (see Sections \ref{hamgeo} and \ref{runtime}).

\section{Construction I: Applying $U$ Once} \label{revisionNagajWocjan}

We begin with the modest goal of applying $U$ to our qubits once ($x=1$) using the time evolution of a simple time-independent and translation-invariant Hamiltonian, where the sequence of gates that implements $U$ is stored in the start state of the qudit chain that we embed our qubits into for the simulation. The construction described in this section only differs from the $d=20$ HQCA construction of Nagaj and Wocjan \cite{NW} in notation (and only slightly at that). We do, however, formalize and elaborate on some of the essential ideas underlying the success and utility of their construction which will be important in subsequent sections when we expand their construction, the most important of which being the definition of what it means for a sequences of quantum states to be `uniquely orthogonally generated' by a sequence of transition rules of an HQCA (see Section \ref{uogsec}). The construction described in this section will be referred to as `construction I' throughout the rest of the paper.

\subsection{Gateset}

The universal gate set we will use to describe a quantum circuit for $U$ is $\mathcal{G}=\{W,S,I\}$ where each gate acts on a local pair of left and right qubits and
\begin{itemize}
	\item $W$ is a controlled rotation by $\pi/2$ about the $y$-axis of the Bloch sphere (where the left qubit always controls the right)
	\item $S$ is the two-qubit swap gate
	\item $I$ is shorthand for the two-qubit identity gate $I\otimes I$. 
\end{itemize}
We note that because all of these gates leave the state $\ket{00}$ invariant, to use these gates to simulate the application of an arbitrary $U$ on a state of the form $\ket{0^n}$, some number of qubits in the $\ket{1}$ state must be provided in order to simulate gates that act non-trivially on the $\ket{0}$ state. This overhead is polynomial in the number of gates that $U$ would have in a circuit using a more traditional gateset, so we ignore it and assume that the state that our gates act on has the correct number of qubits in $\ket{1}$ states for the simulation to proceed.

For the convenience of the construction, the way that $U$ breaks down into a sequence of gates from $\mathcal{G}$ will be as follows: we write the unitary as a gate sequence from $\mathcal{G}$ in the form
\begin{eqnarray}
	U = (U_{K,1}\dots U_{K,N-1})II(U_{K-1,1}\dots U_{K-1,N-1})II \cdots II(U_{1,1}\dots U_{1,N-1})I.
	\label{circuitsequence}
\end{eqnarray}
Here we say that $U$ is expressed as $K$ rounds of $N-1$ gates from $\mathcal{G}$. Note that the use of the term `round' of gates differs here from the usual usage. Normally, a round of gates from some gate set is some tensor product of $m\leq N$ single or two-qubit gates with the property that each of the $N$ qubits has exactly one gate acting on it (identity counted as a gate). This allows the number of rounds making up $U$ to count the time complexity of applying the circuit. This is not true of our rounds. The gate $U_{k,m}\in\mathcal{G}$ is a two qubit gate acting on qubits $m$ and $m+1$. So, two successive gates in a round, $U_{k,m}$ and $U_{k,m+1}$ will possibly each act non-trivially on qubit $m+1$. Thus, the number of rounds, $K$, in our circuit description of $U$ will in general under-count the depth or time complexity of applying $U$ as written, but it will undercount at most by a factor of $N$. The reason we write our circuit this way is for the convenience of constructing our desired HQCA. For ease of discussion, we will use the term \emph{depth} of $U$ to refer to the numbe of rounds, $K$, in the above sense rather than the traditional sense.

Notice also that in the above description, each round of gates is padded with an additional $I$ on either side, except for the final round which does not have an $I$ on the left side. The reason for this is that it makes the construction of our desired HQCA more convenient, and will be explained below. With the above convention for describing $U$ in terms of $\mathcal{G}$, a unitary $U$ of depth $K$ acting on $N$ qubits has a circuit sequence of length $K(N+1)-1$.

\subsection{Hilbert Space}

Here we describe the Hilbert space $\mathcal{C}$ of the 1D chain that our HQCA Hamiltonian $H$ will act on. We imagine our chain as consisting of two registers $\mathcal{C}=P\otimes D$, each with $L$ local sites $\mathcal{C}_j=P_j\otimes D_j$, indexed from left to right. We will describe orthonormal basis states for each local site Hilbert space in terms of symbols:
\begin{itemize}
	\item The program register $P$, whose local sites are $P_j$, stores the sequence of gates that describe $U$ and controls its application to the work qubits. The local hilbert space is 10 dimensional, with basis symbols $\{W,S,I,\overrightarrow{W},\overrightarrow{S},\overrightarrow{I},\gat,\mov,\bul,\rightarrow\}$. So, $P=\mathbb{C}^{10}$.
	\item The data register $D$, whose local sites $D_j$ are qubits described by computational basis symbols $\{0,1\}$, is simply a chain of qubits, $N$ of which are the `work qubits' to which we want to apply the unitary $U$. So, $D=\mathbb{C}^2$. 
\end{itemize}
The symbols in the basis states for $P_j$ have the following interpretation
\begin{eqnarray}
	\wga,\sga,\iga &:& \textrm{representing unitary gates in the program sequence that will apply U,} 
		\nonumber\\
	\rw,\rs,\ri &:& \textrm{marked characters in the program sequence, used to propagate} \nonumber\\
			&& \textrm{the active spot to the front (left) of the program sequence,} \nonumber\\
	\gat &:& \textrm{apply gate symbol,} \nonumber\\
	\mov &:& \textrm{shift program forward,} \nonumber \\
	\rightarrow &:& \textrm{a control symbol indicating that we're in the process of shifting the program } \nonumber \\
	&&\textrm{sequence to the right}, \nonumber \\
	\bul &:& \textrm{empty spot (before/after the program)}. \nonumber
\end{eqnarray}
The way these symbols will interact will become clearer when illustrating the transition rules defined below. The length of the chain will be $L=(2K-1)(N+1)+2$ where $K$ is the number of rounds/depth of the circuit description of $U$, and $N$ is the number of qubits that $U$ is being applied to. The $N$ qubits which $U$ is being applied to will be referred to as the \emph{work qubits}, and their location in the chain $\mathcal{C}$ will be the sites $D_j$ of the data register with $j=(K-1)(N+1)+1+n$ for $n=1...N$.

\subsection{Initial State}

The rest of the construction will be described with a simple illustrative example in mind, where we are applying a $K=2$-round unitary $U=(S_{12}W_{23})II(W_{12}S_{23})I$ to $N=3$ work qubits. Details regarding certain choices about the structure of the initial state will make sense after describing the transition rules and running through applying them to this example initial state. This chain will have $L=14$, and the initial state is written as a product state over the sites and registers: $\ket{\psi_0}=\bigotimes_{j=1}^{14}(\ket{p_j}\otimes\ket{d_j})$ where $\ket{p_j}$ is the state of the $j$th site of the program register and $\ket{d_j}$ is the state of the $j$th site of the data register. The following table describes the initial state:
\begin{eqnarray}
	\begin{array}{c|cccccccccccccc}
		     j & 1 & \cdots     &      &      &      &      & 
		&&&&&& \cdots & L\\
		\hline
		p_j & \rightarrow & S & W & I & I & W &
		S & I & \bul & 
		\bul & \bul & \bul & \bul & \bul 
		\\		
		d_j & 1    & 0    & 0    & 0    & 1    & 
		w_1  & w_2  & w_3  &
		1    & 0    & 0    & 0    & 1	& 0    
 	\end{array} \label{startstate}
\end{eqnarray}

Note the following about the structure of the initial state: 
\begin{itemize}
	\item The individual initial states of the work qubits have here been labelled $w_1$, $w_2$ and $w_3$ to remind the reader where they are located.
	\item The work qubits are bordered by data qubits in the $1$ state on either side
	\item The work qubits are further padded to the left by the sequence $10^N$ $K-1$ times, and to the right by the sequence $0^N1$ $K-1$ times
	\item The final data qubit is in the state $0$
\end{itemize}
The reason for this structure will become clearer in the description and application of the transition rules to this example, and will be commented upon further below.

\subsection{Transition Rules}

We now describe the forward transition rules of our HQCA. Application of these forward transition rules to our initial state $\ket{\psi_0}$ will create a sequence of $T_I+1$ mutually orthogonal states $\ket{\psi_t}_{t=0}^{T_I}$. Note that $T_I$ is a function of $N$ and $K$. The transition rules are deliberately constructed such that for each $\ket{\psi_t}$, there is always one single unique forward transition rule that can be applied to it, which results in the state $\ket{\psi_{t+1}}$. For every forward transition rule there is a unique reverse transition rule (the Hermitian conjugate of the forward transition rule), and for each $\ket{\psi_t}$, there is always one single unique reverse transition rule that can be applied to it, which results in $\ket{\psi_{t-1}}$. No reverse transition rule will be applicable to $\ket{\psi_0}$, and no forward transition rule will be applicable to $\ket{\psi_{T_I}}$.

The transition rules are summarised as follows:
\begin{eqnarray}
	\begin{array}{rccc}
	\raiseonebox 
	1\,:& 
		\band{\rightarrow}{\aga} 
		&\goes&
		\band{\bul}{\overrightarrow{A}} 
		\\
	\end{array}
	\label{rule1}
\end{eqnarray}
\begin{eqnarray}
	\begin{array}{rccc}
	\raiseonebox 
	2\,:& 
		\band{\overrightarrow{A}}{\bga} 
		&\goes&
		\band{\aga}{\rb}
		\\
	\end{array}
	\label{rule2}
\end{eqnarray}
\begin{eqnarray}
	\begin{array}{rccc}
	\raiseonebox 
	3\,:& 
		\band{\overrightarrow{A}}{\bul} 
		&\goes&
		\band{\aga}{\rightarrow} 
		\\
	\end{array}
	\label{rule3}
\end{eqnarray}
\begin{eqnarray}
	\begin{array}{rccccrccc}
	4a\,:&
		\triUL{\rightarrow}{\bul}{1} 
		&\goes&
		\triUL{\gat}{\bul}{1}
	&\quad&
	4b\,:& 
		\triUL{\rightarrow}{\bul}{0} 
		&\goes&
		\triUL{\mov}{\bul}{0}
	\end{array} \label{rule4}
\end{eqnarray}
\begin{eqnarray}
	\begin{array}{rccccrccc}
	5a\,:& 
		\flour{\aga}{\gat}{\psi_{j,j+1}}
		&\goes&
		\flour{\gat}{\aga}{A(\psi_{j,j+1})}
	&\quad&
	5b\,:& 
		\band{\aga}{\mov}
		&\goes&
		\band{\mov}{\aga}
	\end{array}	\label{rule5}
\end{eqnarray}
\begin{eqnarray}
	\begin{array}{rccccrccc}
	6a\,:& 
		\triUL{\bul}{\gat}{1} 
		&\goes&
		\triUL{\bul}{\rightarrow}{1}
	&\quad&
	6b\,:& 
		\triUL{\bul}{\mov}{0} 
		&\goes&
		\triUL{\bul}{\rightarrow}{0}
	\end{array}	\label{rule6}
\end{eqnarray}
Here, $A,~B\in\mathcal{G}$, and $\psi_{j,j+1}$ refers to the state of qubits $j$ and $j+1$ and $A(\psi_{j,j+1})$ refers to the gate $A$ applied to the state of qubits $j$ and $j+1$. We will name this set of forward transition rules $\mathcal{F}_{I}$
\subsection{Active Site}

A subset of the basis states for $P_j$ whose symbols contain directional arrows of some form, $\mathcal{A}_I=\{\rightarrow,\rw,\ri,\rs,\gat,\mov\}$, are called \emph{active symbols}. By construction, the initial state and all states resulting from applying a sequence of transition rules to the initial state will always contain exactly one active symbol, the site at which it is located being referred to as the \emph{active site} of the chain. All transition rules for the HQCA involve the propagation of the active site to a neighbouring site and/or transforming one active symbol into another. This feature will be helpful in analysing the construction below.

\subsection{Illustration}

Now we illustrate the use of these transition rules on our example initial state to demonstrate features of the construction and how the HQCA evolves. We start with initial state $\ket{\psi_0}$ described by the table in equation \ref{startstate}:
\begin{eqnarray}
	\ket{\psi_{0}} &=& \quad \left[ \begin{array}{cccccccccccccc}
		\rightarrow & \sga & \wga & \iga & \iga & 
		\wga & \sga & \iga & \bul &
		\bul & \bul & \bul & \bul & \bul 
		\\		
		1    & 0    & 0    & 0    & 1    & 
		w_1 & w_2 & w_3
		 &
		1    & 0    & 0    & 0    & 1	& 0	
 	\end{array} \right], \label{psi0}
\end{eqnarray}
Applying the sequence of forward transition rules 1, 2 (6 times), and 3, we reach the state,
\begin{eqnarray}
	\ket{\psi_{8}} &=& \quad \left[ \begin{array}{cccccccccccccc}
		\bul & \sga & \wga & \iga & \iga  & 
		\wga & \sga & \iga & \rightarrow  &
		\bul & \bul & \bul & \bul & \bul 
		\\		
		1    & 0    & 0    & 0    & 1    & 
		w_1 & w_2 & w_3
		 &
		1    & 0    & 0    & 0    & 1	& 0	
 	\end{array} \right]. \label{psi8}
\end{eqnarray}
At this point, transition rule 4$a$ applies, turning the $\rightarrow$ symbol into the gate application symbol $\gat$, giving
\begin{eqnarray}
	\ket{\psi_{9}} &=& \quad \left[ \begin{array}{cccccccccccccc}
		\bul & \sga & \wga & \iga & \iga  & 
		\wga & \sga & \iga & \gat &
		\bul & \bul & \bul & \bul & \bul 
		\\		
		1    & 0    & 0    & 0    & 1    & 
		w_1 & w_2 & w_3
		 &
		1    & 0    & 0    & 0    & 1	& 0	
 	\end{array} \right]. \label{psi9}
\end{eqnarray}
Now, rule 5$a$ is applied 3 times, which applies the first round of gates to the work qubits:
\begin{eqnarray}
	\ket{\psi_{12}} &=& \quad \left[ \begin{array}{cccccccccccccc}
		\bul & \sga & \wga & \iga & \iga & \gat & 
		\wga & \sga & \iga & 
		\bul & \bul & \bul & \bul & \bul 
		\\		
		1    & 0    & 0    & 0    & 1    & 
		\multicolumn{3}{c}{S_{23}\left(W_{12}\left(w_1w_2w_3\right)\right)}
		 &
		1    & 0    & 0    & 0    & 1	& 0	
 	\end{array} \right]. \label{psi12}
\end{eqnarray}
Applying rule 5$a$ four more times will have propagated the active site as far back to the left as it can go, giving
\begin{eqnarray}
	\ket{\psi_{16}} &=& \quad \left[ \begin{array}{cccccccccccccc}
		\bul & \gat & \sga & \wga & \iga & \iga & 
		\wga & \sga & \iga & 
		\bul & \bul & \bul & \bul & \bul 
		\\		
		1    & 0    & 0    & 0    & 1    & 
		\multicolumn{3}{c}{S_{23}\left(W_{12}\left(w_1w_2w_3\right)\right)}
		 &
		1    & 0    & 0    & 0    & 1	& 0	
 	\end{array} \right], \label{psi16}
\end{eqnarray}
and applying rule 6$a$ yields
\begin{eqnarray}
	\ket{\psi_{17}} &=& \quad \left[ \begin{array}{cccccccccccccc}
		\bul & \rightarrow & \sga & \wga & \iga & \iga & 
		\wga & \sga & \iga & 
		\bul & \bul & \bul & \bul & \bul 
		\\		
		1    & 0    & 0    & 0    & 1    & 
		\multicolumn{3}{c}{S_{23}\left(W_{12}\left(w_1w_2w_3\right)\right)}
		 &
		1    & 0    & 0    & 0    & 1	& 0	
 	\end{array} \right]. \label{psi17}
\end{eqnarray}


So, we see that applying this \emph{unique} sequence of 17 forward transitions has moved the active site $\rightarrow$ from the far left of the gate sequence over the to the far right of the gate sequence (equation \ref{psi8}), converted the active symbol $\rightarrow$ into the active symbol $\gat$ which is the gate applying symbol (equation \ref{psi9}), which then moved back to the left applying gates to the qubits below, shifting the gate sequence a single site to the right as it went (equations \ref{psi12}, \ref{psi16} and \ref{psi17}). We will refer to this single back and forth movement of the active site as a right-moving \emph{oscillation} of the active site. The halfway point of an oscillation, just before the $\rightarrow$ symbol turns around to move left as $\gat$ (as illustrated in equation \ref{psi8}) is called the \emph{turning point} of the oscillation. 

From the above illustration of the first oscillation, one can now see that applying this same sequence of 17 forward transitions over and over again will simply generate further right moving oscillations of the active site that proceed nearly identically, shifting the entire gate sequence one space to the right with each completed oscillation, until the entire gate sequence has been shifted all the way to the right. For example, after the second right-moving oscillation, the state of the chain will be 
\begin{eqnarray}
	\ket{\psi_{34}} &=& \quad \left[ \begin{array}{cccccccccccccc}
		\bul & \bul & \rightarrow & \sga & \wga & \iga & \iga & 
		\wga & \sga & \iga & 
		\bul & \bul & \bul & \bul
		\\		
		1    & 0    & 0    & 0    & 1    & 
		\multicolumn{3}{c}{S_{23}\left(W_{12}\left(w_1w_2w_3\right)\right)}
		 &
		1    & 0    & 0    & 0    & 1	& 0	
 	\end{array} \right], \label{psi34}
\end{eqnarray}
and after the third oscillation,
\begin{eqnarray}
	\ket{\psi_{51}} &=& \quad \left[ \begin{array}{cccccccccccccc}
		\bul & \bul & \bul & \rightarrow & \sga & \wga & \iga & \iga & 
		\wga & \sga & \iga & 
		\bul & \bul & \bul 
		\\		
		1    & 0    & 0    & 0    & 1    & 
		\multicolumn{3}{c}{S_{23}\left(W_{12}\left(w_1w_2w_3\right)\right)}
		 &
		1    & 0    & 0    & 0    & 1	& 0	
 	\end{array} \right], \label{psi51}
\end{eqnarray}
after the fourth,
\begin{eqnarray}
	\ket{\psi_{68}} &=& \quad \left[ \begin{array}{cccccccccccccc}
		\bul & \bul & \bul & \bul & \rightarrow & \sga & \wga & \iga & \iga & 
		\wga & \sga & \iga & 
		\bul & \bul
		\\		
		1    & 0    & 0    & 0    & 1    & 
		\multicolumn{3}{c}{S_{23}\left(W_{12}\left(w_1w_2w_3\right)\right)}
		 &
		1    & 0    & 0    & 0    & 1	& 0	
 	\end{array} \right], \label{psi51}
\end{eqnarray}
and finally after the fifth,
\begin{eqnarray}
	\ket{\psi_{85}} &=& \quad \left[ \begin{array}{cccccccccccccc}
		\bul & \bul & \bul & \bul & \bul & \rightarrow & \sga & \wga & \iga & \iga & 
		\wga & \sga & \iga & 
		\bul
		\\		
		1    & 0    & 0    & 0    & 1    & 
		\multicolumn{3}{c}{U\left(w_1w_2w_3\right)}
		 &
		1    & 0    & 0    & 0    & 1	& 0	
 	\end{array} \right]. \label{psi51}
\end{eqnarray}
Note that the first round of gates was applied to the work qubits during the first oscillation, no gates were applied during oscillations 2, 3, or 4, and the second round of gates was applied during the fifth oscillation, completing the application of the desired $U$. Whether or not a particular oscillation will apply gates to the data qubits in the second half of the oscillation depends on whether or not there is a 1 or 0 in the data qubit below the active symbol when it reaches the far right of the oscillation. We can see in equations \ref{psi8} and \ref{psi9}, the active symbol $\rightarrow$ becomes an apply gate symbol $\gat$ because there is a $1$ in the data register of the active site at the turning point. If there were a zero instead, which is true at the turning points of the next three oscillations, $\rightarrow$ would turn into $\mov$ and no gates would be applied in the second half of the oscillation. This is good because the second round of gates is not yet aligned with the work qubits (see equations \ref{psi17}, \ref{psi34} and \ref{psi51}), and we do not want to be applying any gates until this alignment is achieved. 

So, we see that the pattern of ones and zeros in the data qubits to the right of the work qubits is simply chosen so that transition rules 4a and 4b can properly enforce that gates are only applied for the first in every $N+1$ oscillations, which is precisely when rounds of gates will properly be aligned to the work qubits that they will be applied to. The pattern of the data qubits to the left of the work qubits mirrors that on the right so that transition rules 6a and 6b can properly convert $\gat$ or $\mov$ back into $\rightarrow$ at the end of the oscillation. Finally, note also that whenever we begin an oscillation that applies gates, rounds of gates that are not being applied during that round are aligned so that they are always applied to $\ket{00}$ qubit states, on which the $S$ and $W$ gates both act trivially. So, we see that the states on the non-work qubits in the data register remain invariant under the application of our HQCA transition rules.

Finally, after one more half oscillation (8 steps) we reach a state $\ket{\psi_{93}}$ to which no forward transition rules apply:
\begin{eqnarray}
	\ket{\psi_{93}} &=& \quad \left[ \begin{array}{cccccccccccccc}
		\bul & \bul &  
		\bul & \bul & \bul & \bul & \sga & \wga & \iga & \iga & 
		\wga & \sga & \iga & \rightarrow
		\\		
		1    & 0    & 0    & 0    & 1    & 
		\multicolumn{3}{c}{U\left(w_1w_2w_3\right)}
		 &
		1    & 0    & 0    & 0    & 1	& 0
 	\end{array} \right]. \label{psi93}
\end{eqnarray}

\subsection{General Case}\label{uogsec}

Now that we understand the construction via the above worked example, we can make some statements about the general case where we are applying a depth-$K$ unitary to $N$ work qubits. In general, if we want to apply depth-$K$ unitary in the form of equation \ref{circuitsequence} to $N$ qubits, we initialize the state of the length $L=(2K-1)(N+1)+2$ chain of $d=20$ qudits as follows:
\begin{itemize}
	\item Program register $P$: \\$P_1=\rightarrow$, $P_2 ... P_{K(N+1)}=(U_{K,1}\dots U_{K,N-1})II(U_{K-1,1}\dots U_{K-1,N-1})II \cdots II(U_{1,1}\dots U_{1,N-1})I$,\\ $P_{j>K(N+1)}=\bul$
	\item Data register $D$: $D_1...D_{(K-1)(N+1)+1}=(1(0)^N)^{K-1}1$, $D_{(K-1)(N+1)+2}...D_{(K-1)(N+1)+1+N}=0^N$ (these are the work qubits), $D_{(K-1)(N+1)+1+N+1}...D_{L=(2K-1)(N+1)+2}=1(0^N1)^{K-1}0$
\end{itemize}

Consider an instance of construction I with a $K$-depth unitary being applied to $N$ work qubits. A single full right-moving oscillation will be achieved through the application of $2K(N+1)+1$ forward transitions: Rule $1$ once and then rule 2 $K(N+1)-2$ times, and then rule 3 once completes the first half of the oscillation where the active symbol moves all the way to the right and gets to the turning point. Then applying rule 4$a$/$b$ once prepares us for the second half of the oscillation where the active site moves all the way back to the left which is accomplished by applying rule 5$a$/$b$ $K(N+1)-1$ times and then applying 6$a$/$b$ once to prepare the active site to begin the next oscillation. The first half of a right-moving oscillation is $K(N+1)$ steps and the second half is $K(N+1)+1$ steps. The final state is reached after $N(K-1)+K$ full oscillations and one half oscillation (of $K(N+1)$ steps). So, the total number of forward transitions needed to reach the end state from the start state is $T_{I}=2N^2K^2-2N^2K+4NK^2-N+2K^2+2K$.

Note that at each step of the sequence of forward transition moves applied on a proper start state $\ket{\psi_0}$, there is only ever exactly one forward transition that can be applied (unless you have reached the end state $\ket{\psi_{T_I}}$). This is, essentially, due to the fact that the start state has a single active symbol, and the transition rules were specifically designed to move and transform the active symbol in a prescribed way.

Nagaj and Wocjan \cite{NW} show that by time evolving the appropriate initial state $\ket{\psi_0}$ encoding the desired unitary under Hamiltonian $H_I$ for time polynomial in $K$ and $N$, a simple measurement on a subset of the chain will collapse the state of the work qubits to $U\ket{0^N}$ with high probability. We defer any run-time analysis to the end of the paper where we will present the run-time analysis for our full construction which is based on this one.

\subsection{Hamiltonian and Hilbert Space Geometry}\label{hamgeo}

Let us choose a Hamiltonian $H_I$ for this system as a sum of translationally invariant terms:
\begin{eqnarray}
	H_I = - \frac{1}{6bL}\sum_{i=1}^{L-1} 
		\sum_{k=1}^{6b} \left( P_{k} + P_{k}^{\dagger} \right)_{(i,i+1)}
		\label{ourHi}
\end{eqnarray}
where the terms ${P_k}_{i,i+1}$ correspond to the rules 1-6b and act on two neighboring qudits as 
\begin{eqnarray}
	{P_1}_{i,i+1} &=& 
		\sum_{A\in\{W,S,I\}}			
			\ket{\bul\overrightarrow{A}}\bra{\rightarrow\aga}_{p_i,p_{i+1}}
		 \otimes 
		\ii_{d_i,d_{i+1}}	
			, \\
	{P_2}_{i,i+1} &=& 
		 \sum_{A,B\in\{W,S,I\}} 
		\ket{\aga\rb}\bra{\overrightarrow{A}\bga}_{p_i,p_{i+1}}
		\otimes
		\ii_{d_i,d_{i+1}} 
		,	 \\
	{P_3}_{i,i+1} &=& 
		 \sum_{A\in\{W,S,I\}}
			\ket{\aga\rightarrow}\bra{\overrightarrow{A}\bul}_{p_i,p_{i+1}}
		\otimes	\ii_{d_i,d_{i+1}}, 
\end{eqnarray}
\begin{eqnarray}
	{P_{4a}}_{i,i+1} &=& 
			\ket{\gat\bul}\bra{\rightarrow\bul}_{p_i,p_{i+1}}
		  \otimes 
			\ket{1}\bra{1}_{d_i}
		  \otimes 
			\ii_{d_{i+1}}, \\
	{P_{4b}}_{i,i+1} &=& 
		\ket{\mov\bul}\bra{\rightarrow\bul}_{p_i,p_{i+1}}
			\otimes
			\ket{0}\bra{0}_{d_i}\otimes \ii_{d_{i+1}}, \\
	{P_{5a}}_{i,i+1} &=& 
		\sum_{A\in\{W,S,I\}}  
			\ket{\gat\aga}\bra{\aga\gat}_{p_i,p_{i+1}}
			\otimes
			A_{d_i,d_{i+1}}, \\
	{P_{5b}}_{i,i+1} &=& 
		\sum_{A\in\{W,S,I\}} \ket{\mov\aga}\bra{\aga\mov}_{p_i,p_{i+1}}
			\otimes
			\ii_{d_i,d_{i+1}}, \\
	{P_{6a}}_{i,i+1} &=& 
		\ket{\bul\rightarrow}\bra{\bul\gat}_{p_i,p_{i+1}}
			\otimes
			\ii_{d_i} \otimes \ket{1}\bra{1}_{d_{i+1}}, \\
	{P_{6b}}_{i,i+1} &=& 
			\ket{\bul\rightarrow}\bra{\bul\mov}_{p_i,p_{i+1}}
			\otimes
			\ii_{d_i} \otimes \ket{0}\bra{0}_{d_{i+1}}.
\end{eqnarray}
Each local term in the sum making up $H$, when acting on a basis state of $\mathcal{C}$, either changes the state by implementing the corresponding HQCA forward/reverse transition rule or annihilates it if the rule can not be applied to that state. The following definition will be essential for further understanding and analyzing the time evolution generated by this Hamiltonian and those based on it later in the paper:
\begin{definition}\label{uog}
	Consider the Hilbert space of states $\mathcal{C}$ of any HQCA described in this paper. A sequence of $T+1$ states from $\mathcal{C}$, $\left\{\ket{\psi_t}\right\}_{t=0}^{T}$, is called \emph{Uniquely Orthogonally Generated} (UOG) by sequence of forward transition rules of the HQCA, $\mathcal{S}$ (which has length $T$), if the two following conditions are satisfied:
	\begin{enumerate}
		\item The states $\left\{\ket{\psi_t}\right\}_{t=0}^{T}$ are each a product state over all sites and registers aside from the work qubits, and are mutually orthogonal (thus the states differ in at least one site of the chain)
		\item For each state $\ket{\psi_{t<T}}$, there is exactly one forward transition rule that can be applied to the state, and the application of this rule yields the state $\ket{\psi_{t+1}}$
	\end{enumerate}
\end{definition}

It is clear from the illustration of a single right-moving oscillation as in the example above that the sequence of intermediate states of the chain throughout a right-moving oscillation is UOG by the sequence of transitions described above.

We will name the unique sequence of forward transitions that takes the initial state of the chain, $\ket{\psi_0}$, to the final state $\ket{\psi_{T_{I}}}$ $\mathcal{T}_{I}\in \mathcal{F}^{T_{I}}_I$. This sequence $\left\{\ket{\psi_t}\right\}_{t=0}^{T_I}$ is UOG by $\mathcal{T}_{I}$. This is easily seen from the fact that for each individual right-moving oscillation, the set of transitions that implements it is unique, and for each successive oscillation the sets of intermediate states are orthogonal to those of all other oscillations because the program sequence starts at a different position in each oscillation. Thus, every state in the sequence $\left\{\ket{\psi_{t}}\right\}_{t=0}^{T_I}$ is orthonormal to all others in the sequence. Further, for each state in $\left\{\ket{\psi_{t}}\right\}_{t=0}^{T_I}$, only one forward transition can be applied at each step because this was true of the individual right-moving oscillations. Thus $\left\{\ket{\psi_t}\right\}_{t=0}^{T_I}$ is UOG by $\mathcal{T}_{I}$.

Now we can see the benefit of the UOG property. The fact that the sequence of states $\left\{\ket{\psi_t}\right\}_{t=0}^{T_I}$ is UOG by $\mathcal{T}_{I}$ allows one to be able to think of the set of states $\ket{\psi_t}$ as the set of positions of a particle on a line, so $H_I$  becomes 
\begin{eqnarray}
	H_{I,line} = - \sum_{t=0}^{T-1} \big(
		\ket{t}\bra{t+1} + \ket{t+1}\bra{t} 
			\big).
	\label{Hwalk}
\end{eqnarray}
This is the Hamiltonian of a (continuous-time) quantum walk on a line of length $T_I+1$. Therefore, the HQCA Hamiltonian $H_I$ induces a continuous quantum walk on the ``line'' of states $\left\{\ket{\psi_{t}}\right\}_{t=0}^{T_I}$ of the qudit chain of length $L$. Note that starting from an appropriately defined start state $\ket{\psi_0}$, the outcome of any time evolution under $H_I$ is restricted to the subspace of states spanned by $\left\{\ket{\psi_{t}}\right\}_{t=0}^{T_I}$.
\subsection{Summary and Conclusions}
\begin{itemize}
	\item Given a depth $K$ unitary on $N$ qubits, we can create a simple initial state $\ket{\psi_0}$ on a qudit chain of length $L=(2K-1)(N+1)+2$ whose local site dimension is $d=20$ (itself factorizing locally into a qubit and a $d=10$ qudit) which encodes the desired unitary
	\item The forward transition rules and their reverse versions define the Hamiltonian $H_{I}$ which encodes an HQCA
	\item Starting from $\ket{\psi_0}$ and applying only forward transition rules, one obtains a \emph{unique} sequence of mutually orthogonal states $\{\ket{\psi_t}\}_{t=0}^{T_I-1}$. For each $\ket{\psi_t}$, there is only ever exactly one forward transition rule that can be applied, from which one obtains $\ket{\psi_{t+1}}$
	\item The continuous time evolution of $\ket{\psi_0}$ under $H_{I}$ can be mapped to a simple 1D quantum walk of a single particle on a chain of length $T_I+1$
	\item The state $\ket{\psi_{T_I}}$ has the desirable feature that it is a product state among all sites and registers except for the work qubits, and the work qubits have had the desired unitary applied to them. No forward transition rule applies to $\ket{\psi_{T_I}}$.
	\item One can do a simple computational basis measurement on the program register to see if the program sequence has moved far enough to the right for $U$ to have been applied to the qubits
\end{itemize}

\section{Construction II - Applying $U$ More Than Once} \label{conii}

What is needed in order to apply $U$ more than once? In the construction that has been made so far, we see that after applying $U$ once to the state of the work qubits, the state of the HQCA reaches a state for which no forward transition rules can be applied. In the language of the quantum walk, the particle has reached the end of the line, so all it can do is turn around and walk the other way, which would undo the application of $U$ to the work qubits. So under arbitrarily long time evolution $U$ will never be applied to the work qubits more than once. To apply $U$ twice, we would desire to be in a state like the start state in Eq. (\ref{psi0}) except with $U$ already applied to the work qubits. So, if there were some way of taking the end state in Eq. (\ref{psi93}) and then somehow move the program sequence back to its starting position \emph{without undoing the application of $U$ to the work qubits}, then we could apply forward transition rules to apply $U$ again. Here, we modify the construction to allow this. We proceed by adding a new set of local basis states to the program register Hilbert states and transition rules that will allow the program sequence to reset itself without undoing the application of $U$. We will refer to the construction described in this section as `construction II'.


\subsection{Hilbert Space}

The following changes are made to the local program register Hilbert space: we add the new basis states represented by symbols $\{\lw,\ls,\li,\rmov,\leftarrow, \tur\}$. The length of the chain is increased by two sites.
\subsection{Initial State}
We must now define a new initial state to account for the fact that we have added two new sites. We will illustrate with the same example $U$ being applied to three work qubits. The example initial state for our new construction will be
\begin{eqnarray}
	\ket{\psi_{0}} &=& \quad \left[ \begin{array}{cccccccccccccccc}
		\tur & \rightarrow & \sga & \wga & \iga & \iga & 
		\wga & \sga & \iga & \bul &
		\bul & \bul & \bul & \bul & \bul & \tur
		\\		
		0 & 1    & 0    & 0    & 0    & 1    & 
		w_1 & w_2 & w_3
		 &
		1    & 0    & 0    & 0    & 1	& 0	& 0
 	\end{array} \right]. \label{psi02}
\end{eqnarray}
The rules for constructing this state are identical to those for the base construction except the two ends of the chain have extra sites added with the program register in state $\tur$ and data register in state $0$.

\subsection{Transition Rules}

We add the following new forward transition rules:
\begin{eqnarray}
	\begin{array}{rccc}
	\raiseonebox 
	7\,:& 
		\band{\aga}{\leftarrow} 
		&\goes&
		\band{\la}{\bul} 
		\\
	\end{array}
	\label{rule7}
\end{eqnarray}
\begin{eqnarray}
	\begin{array}{rccc}
	\raiseonebox 
	8\,:& 
		\band{\bga}{\la} 
		&\goes&
		\band{\lb}{\aga}
		\\
	\end{array}
	\label{rule8}
\end{eqnarray}
\begin{eqnarray}
	\begin{array}{rccc}
	\raiseonebox 
	9\,:& 
		\band{\bul}{\la} 
		&\goes&
		\band{\leftarrow}{\aga} 
		\\
	\end{array}
	\label{rule9}
\end{eqnarray}
\begin{eqnarray}
	\begin{array}{rccc}
	\raiseonebox 
	10\,:& 
		\band{\bul}{\leftarrow} 
		&\goes&
		\band{\bul}{\rmov} 
		\\
	\end{array}
	\label{rule10}
\end{eqnarray}
\begin{eqnarray}
	\begin{array}{rccc}
	\raiseonebox 
	11\,:& 
		\band{\rmov}{\aga} 
		&\goes&
		\band{\aga}{\rmov} 
		\\
	\end{array}
	\label{rule11}
\end{eqnarray}
\begin{eqnarray}
	\begin{array}{rccc}
	\raiseonebox 
	12\,:& 
		\band{\rmov}{\bul} 
		&\goes&
		\band{\leftarrow}{\bul} 
		\\
	\end{array}
	\label{rule12}
\end{eqnarray}
\begin{eqnarray}
	\begin{array}{rccccrccc}
	13a\,:& 
		\band{\rightarrow}{\tur} 
		&\goes&
		\band{\leftarrow}{\tur}
	&\quad&
	13b\,:& 
		\band{\tur}{\leftarrow} 
		&\goes&
		\band{\tur}{\rightarrow}
	\end{array}	\label{rule13}
\end{eqnarray}

The forward transition rules 6-12 are designed to mimic the reverse versions of rules 1-6 using the newly introduced basis symbols. Rule 13 shows conversions between the $\rightarrow$ and $\leftarrow$ symbols via the $\tur$ symbol, called the turn symbol.

\subsection{Illustration}
We now show how the additions made to our construction allow us to repeatedly apply $U$ multiple times. Starting with our new start state
\begin{eqnarray}
	\ket{\psi_{0}} &=& \quad \left[ \begin{array}{cccccccccccccccc}
		\tur & \rightarrow & \sga & \wga & \iga & \iga & 
		\wga & \sga & \iga & \bul &
		\bul & \bul & \bul & \bul & \bul & \tur
		\\		
		0 & 1    & 0    & 0    & 0    & 1    & 
		w_1 & w_2 & w_3
		 &
		1    & 0    & 0    & 0    & 1	& 0	& 0
 	\end{array} \right], \label{psi02}
\end{eqnarray}
the same sequence of forward transitions $\mathcal{T}_{I}$ that took Eq. (\ref{psi0}) to Eq. (\ref{psi93}) yields
\begin{eqnarray}
	\ket{\psi_{T_I}} &=& \quad \left[ \begin{array}{cccccccccccccccc}
		\tur & \bul & \bul &  
		\bul & \bul & \bul & \bul & \sga & \wga & \iga & \iga & 
		\wga & \sga & \iga & \rightarrow & \tur
		\\		
		0 & 1    & 0    & 0    & 0    & 1    & 
		\multicolumn{3}{c}{U\left(w_1w_2w_3\right)}
		 &
		1    & 0    & 0    & 0    & 1	& 0 & 0
 	\end{array} \right]. \label{endstate2}
\end{eqnarray}
Now, whereas in the previous construction, no forward transition rule could apply to $\ket{\psi_{T_I}}$, here there is one forward transition rule that applies to $\ket{\psi_{T_I}}$: rule 13$a$, which brings us to state $\ket{\psi_{T_I+1}}$:
\begin{eqnarray}
	\ket{\psi_{T_I+1}} &=& \quad \left[ \begin{array}{cccccccccccccccc}
		\tur & \bul & \bul &  
		\bul & \bul & \bul & \bul & \sga & \wga & \iga & \iga & 
		\wga & \sga & \iga & \leftarrow & \tur
		\\		
		0 & 1    & 0    & 0    & 0    & 1    & 
		\multicolumn{3}{c}{U\left(w_1w_2w_3\right)}
		 &
		1    & 0    & 0    & 0    & 1	& 0 & 0
 	\end{array} \right]. \label{psitp1}
\end{eqnarray}
Now from state $\ket{\psi_{T_I+1}}$ there is a unique set of $T_I$ forward transitions through mutually orthogonal states that brings us to state $\ket{\psi_{2T_I+1}}$
\begin{eqnarray}
	\ket{\psi_{2T_I+1}} &=& \quad \left[ \begin{array}{cccccccccccccccc}
		\tur & \leftarrow & \sga & \wga & \iga & \iga & 
		\wga & \sga & \iga & \bul &
		\bul & \bul & \bul & \bul & \bul & \tur
		\\		
		0 & 1    & 0    & 0    & 0    & 1    & 
		\multicolumn{3}{c}{U\left(w_1w_2w_3\right)}
		 &
		1    & 0    & 0    & 0    & 1	& 0	& 0
 	\end{array} \right]. \label{psi2tp1}
\end{eqnarray}
The set of transitions $\mathcal{T}_{I}'$ that does this is chosen from rules 7-12. Note that each rule 7-12 is similar to the reverse of one of the rules 1-6. To obtain the set of forward transitions that takes $\ket{\psi_{T_I+1}}\to\ket{\psi_{2T_I+1}}$, take the set of forward transitions that took $\ket{\psi_{0}}\to\ket{\psi_{T_I}}$, $\mathcal{T}_{I}$, reverse it, and replace each transition from 1-6 with its partner from 7-12. Basically, this is undoing everything that happened in the transitions that took $\ket{\psi_{0}}\to\ket{\psi_{T_I}}$ but not undoing the application of gates to the work qubits, because there are no apply gate symbols in the transitions 7-12. It is clear, then that the sequence of states $\left\{ \ket{\psi_{T_I+1+t}} \right\}_{t=0}^{T_I}$ is UOG by $\mathcal{T}_{I}'$ for the same reason that $\left\{ \ket{\psi_{t}} \right\}_{t=0}^{T_I}$ is UOG by $\mathcal{T}_{I}$. From state $\ket{\psi_{2T_I+1}}$, there is only one forward transition rule that applies: 13$b$, which brings us to
\begin{eqnarray}
	\ket{\psi_{2T_I+2}} &=& \quad \left[ \begin{array}{cccccccccccccccc}
		\tur & \rightarrow & \sga & \wga & \iga & \iga & 
		\wga & \sga & \iga & \bul &
		\bul & \bul & \bul & \bul & \bul & \tur
		\\		
		0 & 1    & 0    & 0    & 0    & 1    & 
		\multicolumn{3}{c}{U\left(w_1w_2w_3\right)}
		 &
		1    & 0    & 0    & 0    & 1	& 0	& 0
 	\end{array} \right]. \label{psi2tp2}
\end{eqnarray}
Now, we see that the state is identical in structure to the initial state $\ket{\psi_0}$, but $U$ has already been applied to the work qubits. So, applying the same set of transitions that took $\ket{\psi_0}\to \ket{\psi_{2T_I+2}}$, we should get
\begin{eqnarray}
	\ket{\psi_{4T_I+4}} &=& \quad \left[ \begin{array}{cccccccccccccccc}
		\tur & \rightarrow & \sga & \wga & \iga & \iga & 
		\wga & \sga & \iga & \bul &
		\bul & \bul & \bul & \bul & \bul & \tur
		\\		
		0 & 1    & 0    & 0    & 0    & 1    & 
		\multicolumn{3}{c}{U^2\left(w_1w_2w_3\right)}
		 &
		1    & 0    & 0    & 0    & 1	& 0	& 0
 	\end{array} \right]. \label{psi2tp2}
\end{eqnarray}
This set of forward transitions can be able to be applied over and over again, which will apply $U$ to the work qubits again and again. That is, applying this set of transitions $x$ times should yield the state
\begin{eqnarray}
	\ket{\psi_{x(2T_I+2)}} &=& \quad \left[ \begin{array}{cccccccccccccccc}
		\tur & \rightarrow & \sga & \wga & \iga & \iga & 
		\wga & \sga & \iga & \bul &
		\bul & \bul & \bul & \bul & \bul & \tur
		\\		
		0 & 1    & 0    & 0    & 0    & 1    & 
		\multicolumn{3}{c}{U^x\left(w_1w_2w_3\right)}
		 &
		1    & 0    & 0    & 0    & 1	& 0	& 0
 	\end{array} \right]. \label{psi2tp2}
\end{eqnarray}

\subsection{Active Site}
In the sequence $\mathcal{T}_{I}'$ which resets the gate sequence, the sequence of right moving oscillations that occurred from the sequence $\mathcal{T}_{I}$ play out in reverse, but with the active symbol pointing in the opposite direction this time. Seeing one of the symbols $\mathcal{A}_R=\{\rightarrow,\rw,\ri,\rs,\gat,\mov\}$ at the active site in a state indicates that the system is in the process of applying right moving oscillations that shift the gate sequence to the right, applying the gate sequence to the work qubits when appropriate as it goes, whereas seeing one of the symbols $\mathcal{A}_L=\{\leftarrow,\lw,\li,\ls,\rmov\}$ indicates that the system is in the process of applying left moving oscillations that shift the gate sequence to the left and not applying any gates to the work qubits.

\subsection{Conclusion: There is A Catch}

We see that we have built a sequence of states $\left\{\ket{\psi_t}\right\}_{t=0}^{T_{II}=2T_I+2}$, for which each state has a unique forward transition described by the sequence of transitions $\mathcal{T}_{II}=\{\mathcal{T}_I,13a,\mathcal{T}_I',13b\}$, and for which $\ket{\psi_{T_{II}}}$ is identical to $\ket{\psi_0}$ except that $U$ has been applied to the work qubits. We see, then, that by applying the set of transitions $\mathcal{T}_{II}$ repeatedly, the work qubits have $U$ applied to them repeatedly.

The above conclusion would tempt one to say that by building a new Hamiltonian $H_{II}$ analogous to $H_{I}$ but with the new additional transition rules, we again have a system whose time evolution is analogous to a 1D single particle quantum walk where the farther the particle walks from the starting point of the chain $\ket{\psi_0}$, the more times $U$ will be applied to the work qubits. This, however, is incorrect. For the states $\{\ket{\psi_t}\}_{t=0}^{xT_{II}}$ to map to adjacent position eigenstates of a particle on a chain to, the states must be UOG by $\mathcal{T}_{II}^x$. However, they are not. The only difference in state $\ket{\psi_t}$ and state $\ket{\psi_{t+kT_{II}}}$ is the number of times that $U$ has been applied to the work qubits. For any quantum state $\ket{\phi}$, $\ket{\phi}$ and $U^k\ket{\phi}$ will not generally be orthogonal.

As a result, if we time evolve the state $\ket{\psi_0}$ with $H_{II}$, states that have $U$ applied to the work qubits a different number of times can interfere with each other, and can not be distinguished by a simple measurement. There is no clear way to do a measurement to collapse the time evolved state into one where $U$ has been applied a certain number of times. We will remedy this problem by further augmenting our construction to include a binary clock that `counts' the number of times that $U$ has been applied and orthogonalizes the states which have had $U$ applied a different number of times.

\section{Construction III: Binary Clock Construction}\label{coniii}

We would like to modify Construction II so that by applying forward transition rules, all of the uniquely obtained states of the chain at each step, $\ket{\psi_t}$, are mutually orthogonal, and as a particular consequence, states of the chain where the work qubits have had $U$ applied a different number of times will be orthogonal, and can be distinguished by doing a computational basis state measurement on a subset of the chain that excludes the work qubits. We will accomplish this by further modifying the construction in the previous section. First, we give an overview of how the clock itself should work, then we describe how to implement its control and integration into our existing construction, and then we will prove its correctness. Similar binary clock constructions appear in \cite{AGIK,GI,CPGW}.

\subsection{Hilbert Space}

The clock, itself, will consist of two registers/layers: the clock register (denoted $C$), whose local sites $C_j$ are four dimensional with basis $\{\bul,X,0,1\}$, and the clock pointer register (denoted $CP$), whose local sites $CP_j$ are five dimensional with basis states $\{\bul,X,L,R,C\}$. So we add these two new registers to our system ($C$ being the third layer, $CP$ being the fourth). We will also augment the program register's local Hilbert space basis to include the symbol $\downdownarrows$.

So, our complete local Hilbert space description is as follows: There are $L=(2K-1)(N+1)+4$ local sites on the chain, each site $j$ consisting of four different local Hilbert spaces, one from each register:
\begin{itemize}
	\item The program register's local sites $P_j$ have basis $\{W,S,I,\overrightarrow{W},\overrightarrow{S},\overrightarrow{I},\gat,\mov,\bul,\rightarrow,\tur,\li,\lw,\ls,\rmov,\leftarrow,\downdownarrows\}$
	\item The data register's local sites $D_j$ have basis $\{0,1\}$
	\item The clock register's local sites $C_j$ have basis $\{\bul,0,1\}$
	\item The clock pointer register's local sites $CP_j$ have basis $\{\bul,X,L,R,C\}$
\end{itemize}
The total local site dimension is $16\times 2\times 3\times 5=480$.

\subsection{Initial State}

The initial state for this augmented construction is as follows:
\begin{eqnarray}
	\ket{\psi_{0}} &=& \quad \left[ \begin{array}{cccccccccccccccc}
		\tur & \bul &
		\bul & \bul & \bul & \bul & \bul & \sga & \wga & \iga & \iga & 
		\wga & \sga & \iga & \tur & \tur
		\\		
		0 & 1    & 0    & 0    & 0    & 1    & 
		w_1 & w_2 & w_3
		 &
		1    & 0    & 0    & 0    & 1	& 0	& 0
		\\
		\bul & 0 & 1 & 1 & 1 & 1 & 1 & 1 & 1 & 1 & 1 & 1 & 1 & 1 & 1 & 1
		\\
		\bul & R & \bul & \bul & \bul & \bul & \bul & \bul & \bul & \bul & \bul & \bul & \bul & \bul & \bul & \bul
 	\end{array} \right], \label{psi03}
\end{eqnarray}

Here, the third layer is the clock register and the fourth and final layer is the clock pointer register. Notice that the program register has a different initial configuration than the previous two constructions: this time, the gate sequence starts shifted all the way to the right, and the clock and clock pointers are specifically initialized so that no reverse transition rule applies to the initial state. This will be illustrated below.

\subsection{Clock Schematic and Update Algorithm}

Here we describe a simple algorithm for updating a binary counter which will be implemented in our HQCA transition rules. The idea is the following: The clock register, consisting of $L$ sites, can store a binary number of up to $L-1$ digits long. Numbers with $n<L$ significant digits will be stored in the $n$ rightmost sites of the clock register via the $0$ and $1$ states, and the remaining $L-n$ sites will all be in the state $0$ except for the first register which will always be $\bul$. Significant bits will increase from right to left, so, for example, the binary representation of the number 6, in our clock, will look like $\bul 0...00000110$.

Given any binary number represented in this form, we can use the following schematic algorithm for incrementing it by $1$:
\begin{itemize}
	\item A) Initialize a `pointer' which starts sitting underneath the least significant bit.
	\item B) If the least significant bit is 0, change it to a 1, end
	\item C) If the least significant bit is 1, and the next bit is 0, update both bits $10\to01$, end
	\item D) Otherwise, the first two bits are $11$. In this case, move the pointer to the left, bit by bit, until it encounters a $1$ bit whose next most significant bit is a $0$. When it is found, update these two bits $01\to10$. Move the pointer back to the right, bit by bit, flipping all of the less significant $1$ bits to $0$ as it passes, until it is back under the least significant bit, end.
	\item E) If the number being incremented is $\bul 11... 1$, then the pointer halts when it reaches the most significant $1$, as this number can not be incremented further.
\end{itemize}
In our construction, the $CP$ register is where the pointer that scans through and increments the clock registers will live. Generally, when the clock is not being incremented, the $CP$ register will be in the state $\bul\bul...\bul X$. This indicates that the `pointer' is underneath the least significant bit of the clock, and is inactive, $X$. The $\bul$ symbol underneath all of the other bits simply indicates the absence of the clock pointer. When the clock pointer is scanning to the left, looking for a $1$ bit whose next most significant bit is a $0$, it will be in state $L$ and will hop right past the $\bul$ symbols until it encounters this configuration in the clock register. Then it updates the clock bits according to the algorithm above, the pointer in state $L$ transitions to a pointer in state $R$, and will now start scanning back to the right, flipping the less significant $1$ bits it encounters along the way to $0$ bits. Once it reaches the least significant bit, the clock pointer transitions to the $C$ state to indicate that the clock update is complete, and then back to the $X$ state, in which it will stay until it is once again called upon to increment the clock. 
\subsection{Transition Rules}
Now we give the transition rules which implement the control of the clock as described above. These rules include a modification to rule 13:
\begin{eqnarray}
	\begin{array}{rccccrccc}
	13a\,:& 
		\band{\rightarrow}{\tur}
		&\goes&
		\band{\tur}{\downdownarrows}
	&\quad&
	13b\,:& 
		\band{\tur}{\leftarrow}
		&\goes&
		\band{\tur}{\rightarrow}
	\end{array}	\label{rule13}
\end{eqnarray}
\begin{eqnarray}
	\begin{array}{rccccrccc}
	14\,:& 
		\eight{\tur}{\downdownarrows}{-}{-}{-}{-}{\bul}{X}
		&\goes&
		\eight{\tur}{\tur}{-}{-}{-}{-}{\bul}{L}
	\end{array}	\label{rule14}
\end{eqnarray}
\begin{eqnarray}
	\begin{array}{rccccrccc}
	15\,:& 
		\eight{\tur}{\tur}{-}{-}{-}{0}{\bul}{L}
		&\goes&
		\eight{\tur}{\tur}{-}{-}{-}{1}{\bul}{C}
	\end{array}	\label{rule15}
\end{eqnarray}
\begin{eqnarray}
	\begin{array}{rccccrccc}
	16\,:& 
		\eight{\tur}{\tur}{-}{-}{0}{1}{\bul}{L}
		&\goes&
		\eight{\tur}{\tur}{-}{-}{1}{0}{\bul}{C}
	\end{array}	\label{rule16}
\end{eqnarray}
\begin{eqnarray}
	\begin{array}{rccccrccc}
	17\,:& 
		\eight{-}{-}{-}{-}{1}{1}{\bul}{L}
		&\goes&
		\eight{-}{-}{-}{-}{1}{1}{L}{\bul}
	\end{array}	\label{rule17}
\end{eqnarray}
\begin{eqnarray}
	\begin{array}{rccccrccc}
	18\,:& 
		\eight{not~\tur}{-}{-}{-}{0}{1}{\bul}{L}
		&\goes&
		\eight{not~\tur}{-}{-}{-}{1}{0}{\bul}{R}
	\end{array}	\label{rule18}
\end{eqnarray}
\begin{eqnarray}
	\begin{array}{rccccrccc}
	19\,:& 
		\eight{not~\tur}{-}{-}{-}{0}{1}{R}{\bul}
		&\goes&
		\eight{not~\tur}{-}{-}{-}{0}{0}{\bul}{R}
	\end{array}	\label{rule19}
\end{eqnarray}
\begin{eqnarray}
	\begin{array}{rccccrccc}
	20\,:& 
		\eight{\tur}{\tur}{-}{-}{0}{1}{R}{\bul}
		&\goes&
		\eight{\tur}{\tur}{-}{-}{0}{0}{\bul}{C}
	\end{array}	\label{rule20}
\end{eqnarray}
\begin{eqnarray}
	\begin{array}{rccccrccc}
	21\,:& 
		\eight{\tur}{\tur}{-}{-}{-}{-}{\bul}{C}
		&\goes&
		\eight{\leftarrow}{\tur}{-}{-}{-}{-}{\bul}{X}
	\end{array}	\label{rule21}
\end{eqnarray}
\subsection{Active Site}
In analyzing the correctness of the above transition rules for achieving our goals, it will again be useful to make use of the concept of the active site and active symbol. In our augmented construction, there will again only ever be exactly one active symbol in the state, which will indicate the active site. But this time, the active symbol can either be in the program register \emph{or} the clock pointer register, depending on whether or not the clock is being updated. The set of active symbols is now the union of the active symbols for the $P$ and $CP$ registers: $\mathcal{A}=\mathcal{A}_P\cup\mathcal{A}_{CP}$, with $\mathcal{A}_P=\{\leftarrow,\rightarrow,\lw,\ls,\li,\rw,\rs,\ri,\mov,\gat,\rmov,\downdownarrows\}$ and $\mathcal{A}_{CP}=\{L,R,C\}$. Transition rules 14 and 21 describe the passing of the active symbol between the program and clock pointer registers. It is still true that every single transition rule describes the moving of an active symbol from one site to a neighbouring site and/or the conversion of one kind of active symbol into another, which is essential for our sequence of states to be UOG and for the dynamics of the HQCA to map to a quantum walk on a line.

\subsection{Correctness}
We claim the following: starting from the initial state
\begin{eqnarray}
	\ket{\psi_{0}} &=& \quad \left[ \begin{array}{cccccccccccccccc}
		\tur & \bul &
		\bul & \bul & \bul & \bul & \bul & \sga & \wga & \iga & \iga & 
		\wga & \sga & \iga & \tur & \tur
		\\		
		0 & 1    & 0    & 0    & 0    & 1    & 
		w_1 & w_2 & w_3
		 &
		1    & 0    & 0    & 0    & 1	& 0	& 0
		\\
		\bul & 0 & 1 & 1 & 1 & 1 & 1 & 1 & 1 & 1 & 1 & 1 & 1 & 1 & 1 & 1
		\\
		\bul & R & \bul & \bul & \bul & \bul & \bul & \bul & \bul & \bul & \bul & \bul & \bul & \bul & \bul & \bul
 	\end{array} \right], \label{psi03}
\end{eqnarray}
and applying forward transition rules 1-21, one generates a unique and finite sequence of mutually orthogonal states $\left\{\ket{\psi_t}\right\}_{t\geq 0}$, whose states on all registers across all layers aside from the work qubits are orthogonal product states. Moreover,
\\
\\
A) The final state is
\begin{eqnarray}
	\ket{\psi_{F}} &=& \quad \left[ \begin{array}{cccccccccccccccc}
		\tur &\bul & \bul & \bul & \bul & \bul & \bul & \sga & \wga & \iga & \iga & 
		\wga & \sga & \iga & \tur & \tur
		\\		
		0 & 1    & 0    & 0    & 0    & 1    & 
		\multicolumn{3}{c}{U^k\left(w_1w_2w_3\right)}
		 &
		1    & 0    & 0    & 0    & 1	& 0	& 0
		\\
		\bul & 1 & 1 & 1 & 1 & 1 & 1 & 1 & 1 & 1 & 1 & 1 & 1 & 1 & 1 & 1
		\\
		\bul & L & \bul & \bul & \bul & \bul & \bul & \bul & \bul & \bul & \bul & \bul & \bul & \bul & \bul & \bul
 	\end{array} \right], \label{psi03}
\end{eqnarray}
where $U$ has been applied to the work qubits $k=2^{L-1}$ times, where $L=(2K-1)(N+1)+4$ is the number of sites in the chain.
\\
\\
B) Suppose you define a Hamiltonian, $H_{III}$, out of terms that implement transition rules 1-21 and their Hermitian conjugates, and use this Hamiltonian to continuously time evolve the state $\ket{\psi_0}$. If, at any point in the time evolution, one measures the clock layer to hold the binary number $k$, and the first site of the $CP$ layer to hold the state $C$, then the post measurement state of the work qubits is guaranteed to be in the state $U^k\ket{w_1 w_2 w_3}$.
\\
\\
I will separate the proof into several small pieces.
\begin{lem1}
	Starting from the state
	\begin{eqnarray}
		\ket{S^{U,x}_0} &=& \quad \left[ \begin{array}{cccccccccccccccc}
			\tur & \bul & 
			\bul & \bul & \bul & \bul & \bul & \sga & \wga & \iga & \iga & 
		\wga & \sga & \iga & \leftarrow & \tur
			\\		
			0    & 1    & 0    & 0    & 0    & 1    & 
			\multicolumn{3}{c}{U^{x}\left(w_1w_2w_3\right)}  &
			1    & 0    & 0    & 0    & 1    & 0 & 0
			\\
			\bul & \multicolumn{15}{c}{binary~x}
			\\
			\bul & \bul & \bul & \bul & \bul & \bul & \bul & \bul & \bul & \bul & \bul & \bul & \bul & \bul & \bul & X
	 	\end{array} \right]. \label{endstate}
	\end{eqnarray}
	there is a sequence of states $S^{U,x}=\left\{\ket{S^{U,x}_t}\right\}_{t= 0}^{n_{U}}$ UOG by sequence of $n_U$ forward transitions $\mathcal{U}$ and whose final state is
	\begin{eqnarray}
			\ket{S^{U,x}_{n_U}} &=& \quad \left[ \begin{array}{cccccccccccccccc}
				\tur & \bul & 
				\bul & \bul & \bul & \bul & \bul & \sga & \wga & \iga & \iga & 
			\wga & \sga & \iga & \rightarrow & \tur
				\\		
				0    & 1    & 0    & 0    & 0    & 1    & 
				\multicolumn{3}{c}{U^{x+1}\left(w_1w_2w_3\right)}  &
				1    & 0    & 0    & 0    & 1    & 0 & 0
				\\
				\bul & \multicolumn{15}{c}{binary~x}
				\\
				\bul & \bul & \bul & \bul & \bul & \bul & \bul & \bul & \bul & \bul & \bul & \bul & \bul & \bul & \bul & X
		 	\end{array} \right]. \label{endstate}
		\end{eqnarray}
	for any $x\in\{0,1,\cdots,2^{L-1}-1\}$.
\end{lem1}
\begin{proof}
By the correctness of Construction II, the sequence of forward transitions $\mathcal{U}=\{\mathcal{T}_I',13b,\mathcal{T}_I\}$ fulfills the requirements.
\end{proof}
The sequence of forward transitions $\mathcal{U}$ described above will referred to as an \emph{application transition} since it generates a single application of $U$ to the work qubits, and $n_U=T_{II}-1$ is the number of transitions in $\mathcal{U}$. The sequence of mutually orthogonal states $S^{U,x}=\left\{ \ket{S_t^{U,x}} \right\}_{t=0}^{n_U}$ is called the $x$th application sequence.
\begin{lem2}
Starting with a state of the form 
	\begin{eqnarray}
			\ket{S^{C,x}_0} &=& \quad \left[ \begin{array}{cccccccccccccccc}
				\tur & \bul & 
				\bul & \bul & \bul & \bul & \bul & \sga & \wga & \iga & \iga & 
			\wga & \sga & \iga & \tur & \tur
				\\		
				0    & 1    & 0    & 0    & 0    & 1    & 
				\multicolumn{3}{c}{U^{x+1}\left(w_1w_2w_3\right)}  &
				1    & 0    & 0    & 0    & 1    & 0 & 0
				\\
				\bul & \multicolumn{15}{c}{binary~x}
				\\
				\bul & \bul & \bul & \bul & \bul & \bul & \bul & \bul & \bul & \bul & \bul & \bul & \bul & \bul & \bul & L
		 	\end{array} \right], \label{endstate}
		\end{eqnarray}
the sequence of states $S^{C,x}=\left\{\ket{S^{C,x}_t}\right\}_{t= 0}^{n_{C,x}}$ is UOG by sequence of forward transitions $\mathcal{CU}_{x}$, with final state
	\begin{eqnarray}
			\ket{S^{C,x}_{n_{C,x}}} &=& \quad \left[ \begin{array}{cccccccccccccccc}
				\tur & \bul & 
				\bul & \bul & \bul & \bul & \bul & \sga & \wga & \iga & \iga & 
			\wga & \sga & \iga & \tur & \tur
				\\		
				0    & 1    & 0    & 0    & 0    & 1    & 
				\multicolumn{3}{c}{U^{x+1}\left(w_1w_2w_3\right)}  &
				1    & 0    & 0    & 0    & 1    & 0 & 0
				\\
				\bul & \multicolumn{15}{c}{binary~x+1}
				\\
				\bul & \bul & \bul & \bul & \bul & \bul & \bul & \bul & \bul & \bul & \bul & \bul & \bul & \bul & \bul & C
		 	\end{array} \right]. \label{endstate}
		\end{eqnarray}
for all $x\in\{0,1,\cdots,2^{L-1}\}$.
\end{lem2}
\begin{proof}
Once a state of the form $\ket{S^{C,x}_0}$ is reached, there is no longer any active symbol in the program register, and the active symbol is now in the CP register.

First, observe that no matter what number the clock layer holds, there will only ever be one of four possibly valid forward transitions from a state of the form $\ket{S^{C,k}_0}$, which will be either rule 15, 16, or 17 depending on the values of the clock's least two significant bits. Referring to the clock update algorithm described earlier, rule 15 corresponds to situation B), rule 16 corresponds to situation C), and rule 17 corresponds to situation D).

If the least significant bit of the clock in state $\ket{S^{C,x}_0}$ is zero (least two significant bits are $10$ or $00$), then only rule 15 can be applied, resulting in state 
	\begin{eqnarray}
			\ket{S^{C,x}_{n_{C,x}}} &=& \quad \left[ \begin{array}{cccccccccccccccc}
				\tur & \bul & 
				\bul & \bul & \bul & \bul & \bul & \sga & \wga & \iga & \iga & 
			\wga & \sga & \iga & \tur & \tur
				\\		
				0    & 1    & 0    & 0    & 0    & 1    & 
				\multicolumn{3}{c}{U^{x+1}\left(w_1w_2w_3\right)}  &
				1    & 0    & 0    & 0    & 1    & 0 & 0
				\\
				\bul & \multicolumn{15}{c}{binary~x+1}
				\\
				\bul & \bul & \bul & \bul & \bul & \bul & \bul & \bul & \bul & \bul & \bul & \bul & \bul & \bul & \bul & C
		 	\end{array} \right]. \label{endstate}
		\end{eqnarray}
so we're done. (By examining transition rule $15$, it's obvious that in any of these situations, applying the single transition rule correctly increments the clock's stored number, as per case $B$ of the clock update algorithm described earlier.)

If the two least significant bits of the clock in state $\ket{S^{C,x}_0}$ are $01$, then only rule 16 can be applied, resulting in state 
	\begin{eqnarray}
			\ket{S^{C,x}_{n_{C,x}}} &=& \quad \left[ \begin{array}{cccccccccccccccc}
				\tur & \bul & 
				\bul & \bul & \bul & \bul & \bul & \sga & \wga & \iga & \iga & 
			\wga & \sga & \iga & \tur & \tur
				\\		
				0    & 1    & 0    & 0    & 0    & 1    & 
				\multicolumn{3}{c}{U^{x+1}\left(w_1w_2w_3\right)}  &
				1    & 0    & 0    & 0    & 1    & 0 & 0
				\\
				\bul & \multicolumn{15}{c}{binary~x+1}
				\\
				\bul & \bul & \bul & \bul & \bul & \bul & \bul & \bul & \bul & \bul & \bul & \bul & \bul & \bul & \bul & C
		 	\end{array} \right]. \label{endstate}
		\end{eqnarray}
so we're done. (Again, by examining transition rule $16$, it's obvious that in any of these situations, applying the single transition rule correctly increments the clock's stored number, as per case $C$ of the clock update algorithm described earlier.)

If rules 15 and 16 don't apply to $\ket{S^{C,x}_0}$, then let $2\leq p
\leq\left\lceil{\log_{2}(x)}\right\rceil$ be the number of least significant bits of $x$ that form an uninterrupted string of $1$s. Then it is clear that there are $p-1$ unique forward transitions through $p-1$ applications of rule 17. Because $x<2^{L-1}$, we know that $p< L-1$. Then, after the $p-1$ applications of rule 17, registers $p$ and $p+1$ of the chain must fit the description of the initial state of rule 18, in which case rules 15-17 can not apply (after the first application of 17, the program registers won't have the $\tur$'s needed in the program register to allow 15 or 16 to apply), therefore only rule 18 can be applied because it is the only remaining transition rule whose initial state has $L$ in the clock pointer.

After rule 18 is applied, the clock pointer's active site is in $R$ mode, so only rules 19 or 20 could potentially be applied. If $p=2$, then only rule 20 can be applied because the existence of the $\tur$ symbol in the program register above the $R$ symbol will forbid application of rule 19. If $p>2$, then after application of rule 18, we see that bits $p$ and $p-1$ of the number stored in the clock (counting from the right) are exactly the requisite initial state for applying rule 19, so 19 is the only way forward. This will obviously be true for $p-2$ applications of rule 19, at which point the $R$ symbol will have the $\tur$ symbol in the program register above it, so only rule 20 will be able to be applied. In either case, at this point we will have implemented case E) of the clock increase algorithm, and will be in state $\ket{S^{C,x}_{n_{C,x}}}$ as above. Note that the state of the program and data layers never changed at any point in this sequence because we only used rules $15$ through $20$, and they leave those layers invariant.

By having described all possibilities, we see that the sequence $S^{C,x}$ is UOG by the appropriate sequence of forward transitions $\mathcal{CU}_{x}$.
\end{proof}
The sequence of forward transitions, $\mathcal{CU}_{x}$, that takes $\ket{S^{C,x}_0}\to\ket{S^{C,x}_{n_{C,x}}}$ is called a \emph{clock transition}, and the sequence of mutually orthogonal states $S^{C,x}=S^{C,x}=\left\{\ket{S^{C,x}_t}\right\}_{t= 0}^{n_{C,x}}$ is called the $x$th clock sequence.
\begin{lem3}
The states in sequences $S^{C,k}$ and $S^{C,j}$ are mutually orthogonal for $j\neq k$.
\end{lem3}
\begin{proof}
Suppose that there were states $\ket{\phi_j}\in S^{C,j}$ and $\ket{\phi_k}\in S^{C,k}$ which are not orthogonal. This necessarily means that the configurations of the clock and clock pointer registers in both states are identical. This, then, implies that it is possible to find a sequence of forward transitions to apply to $\ket{\phi_j}$ that brings it to $\ket{S^{C,k}_{n_{C,k}}}$. But, then, this contradicts Lemma 2 which states that the clock update sequence $S^{C,k}$ with is UOG, as this would imply that there are at least two distinct forward transition rules that could be applied to $\ket{\phi_j}$.
\end{proof}
\begin{lem4}
The states in sequences $S^{P,k}$ and $S^{P,j}$ are mutually orthogonal for $j\neq k$.
\end{lem4}
\begin{proof}
This is obvious because the clocks will hold different numbers in each sequence ($j$ and $k$ respectively).
\end{proof}
\begin{lem5}
Starting from the state
		\begin{eqnarray}
			\ket{S^{C,2^{L-1}-1}_{0}} &=& \quad \left[ \begin{array}{cccccccccccccccc}
				\tur & \bul & 
				\bul & \bul & \bul & \bul & \bul & \sga & \wga & \iga & \iga & 
			\wga & \sga & \iga & \tur & \tur
				\\		
				0    & 1    & 0    & 0    & 0    & 1    & 
				\multicolumn{3}{c}{U^{2^{L-1}}\left(w_1w_2w_3\right)}  &
				1    & 0    & 0    & 0    & 1    & 0 & 0
				\\
				\bul & 1 & 1& 1& 1& 1& 1& 1& 1& 1& 1& 1& 1& 1& 1& 1
				\\
				\bul & \bul & \bul & \bul & \bul & \bul & \bul & \bul & \bul & \bul & \bul & \bul & \bul & \bul & \bul & L
		 	\end{array} \right]. \label{endstate}
		\end{eqnarray}
there is a unique sequence of $L-2$ forward transitions which maps us to the final state
	\begin{eqnarray}
		\ket{S^{C,2^{L-1}-1}_{0}} &=& \quad \left[ \begin{array}{cccccccccccccccc}
			\tur & \bul & 
			\bul & \bul & \bul & \bul & \bul & \sga & \wga & \iga & \iga & 
		\wga & \sga & \iga & \tur & \tur
			\\		
			0    & 1    & 0    & 0    & 0    & 1    & 
			\multicolumn{3}{c}{U^{2^{L-1}}\left(w_1w_2w_3\right)}  &
			1    & 0    & 0    & 0    & 1    & 0 & 0
			\\
			\bul & 1 & 1& 1& 1& 1& 1& 1& 1& 1& 1& 1& 1& 1& 1& 1
			\\
			\bul & L & \bul & \bul & \bul & \bul & \bul & \bul & \bul & \bul & \bul & \bul & \bul & \bul & \bul & \bul
	 	\end{array} \right]. \label{endstate}
	\end{eqnarray}
from which there are no possible forward transitions.
\end{lem5}
\begin{proof}
Clearly, applying rule 17 $L-2$ times does the job.
\end{proof}
\begin{lem6}
Starting from the initial state
\begin{eqnarray}
	\ket{\psi_{0}} &=& \quad \left[ \begin{array}{cccccccccccccccc}
		\tur & \bul &
		\bul & \bul & \bul & \bul & \bul & \sga & \wga & \iga & \iga & 
		\wga & \sga & \iga & \tur & \tur
		\\		
		0 & 1    & 0    & 0    & 0    & 1    & 
		w_1 & w_2 & w_3
		 &
		1    & 0    & 0    & 0    & 1	& 0	& 0
		\\
		\bul & 0 & 1 & 1 & 1 & 1 & 1 & 1 & 1 & 1 & 1 & 1 & 1 & 1 & 1 & 1
		\\
		\bul & R & \bul & \bul & \bul & \bul & \bul & \bul & \bul & \bul & \bul & \bul & \bul & \bul & \bul & \bul
 	\end{array} \right], \label{psi03}
\end{eqnarray}
and applying forward transition rules 1-21, one generates the sequence $\{\ket{\psi_t}\}_{t=0}^F$ which is UOG by some sequence of $F$ transitions, $\mathcal{S}$, and whose final state is reaching the final state
	\begin{eqnarray}
		\ket{\psi_F} &=& \quad \left[ \begin{array}{cccccccccccccccc}
			\tur & \bul & 
			\bul & \bul & \bul & \bul & \bul & \sga & \wga & \iga & \iga & 
		\wga & \sga & \iga & \tur & \tur
			\\		
			0    & 1    & 0    & 0    & 0    & 1    & 
			\multicolumn{3}{c}{U^{2^{L-1}}\left(w_1w_2w_3\right)}  &
			1    & 0    & 0    & 0    & 1    & 0 & 0
			\\
			\bul & 1 & 1& 1& 1& 1& 1& 1& 1& 1& 1& 1& 1& 1& 1& 1
			\\
			\bul & L & \bul & \bul & \bul & \bul & \bul & \bul & \bul & \bul & \bul & \bul & \bul & \bul & \bul & \bul
	 	\end{array} \right]. \label{endstate}
	\end{eqnarray}
\end{lem6}
\begin{proof}
By lemmas $1$ through $5$, we see that the sequence of forward transitions of length $F$, $\mathcal{S}=\{\mathcal{L}_{-1},\{\mathcal{L}_x\}_{x=0}^{2^{L-1}-1},\mathcal{L}_F\}$, with subsequences defined
\begin{eqnarray}
	\mathcal{L}_{-1} &=& \{ 19^{L-3},20,21 \}
	\\
	\mathcal{L}_x&=&\{\mathcal{U},13a,14,\mathcal{UC}_x,21\}
	\\
	\mathcal{L}_F&=&\{\mathcal{U},13a,14,17^{L-2}\}
\end{eqnarray}
does the job.
\end{proof}

\section{Construction IV: Deterministic Selection of $x$ in time {\rm{poly}}($x,N$)}\label{coniv}
Now we augment construction III so that a particular number $x$ of applications of $U$ can be selected to be applied to the work qubits, and we can obtain the state where this has happened more or less deterministically by time evolving for a time polynomial in $x$ and $N$. To do this, we build in a new register that holds the number of desired applications $x$ in binary form, and once the number in the clock register reaches $x$, forward transitions are modified so that although the quantum walk continues forward, gates are no longer applied to data qubits, so there is some $T<F$ such that for all $\ket{\psi_{t>T}}$, the work qubits will always have $U^x$ applied to them.
\subsection{Hilbert Space}
We add two new registers in this construction: the target register $T$, and the a second clock register $C_2$. The local sites of both registers are three dimensional $\{0,1,\bul\}$. We also make the following additions of new states to the local sites of the other registers:
\begin{itemize}
	\item The program register adds symbols $\{\overrightarrow{I}^{\times},\overrightarrow{S}^{\times},\overrightarrow{W}^{\times},\mov^{\times},\rightarrow^{\times},\li^{\times},\ls^{\times},\lw^{\times},\rmov^{\times},\leftarrow^{\times},\downdownarrows^{\times}\}$
	\item The clock pointer register adds symbols $\{\overleftarrow{C},CX, R^{\times},C^{\times}, L^{\times}\}$
\end{itemize}
With these additions, the local dimensions of the program, data, clock, clock pointer, target and second clock pointer registers are now 27, 2, 3, 10, 3, and 3 respectively. The total local dimension of a site of the qudit chain $\mathcal{C}$ is thus $d=27\times 2\times 3 \times 10\times 3\times 3=14580$.
\subsection{Initial State}
In the initial state we must now have initialization of the fifth and sixth layers, the target and second clock registers respectively. The target register is initialized to hold the binary representation of the target number of applications, $x$, with significant digits increasing from right to left, and all places greater than the most significant $1$ of $x$ being padded by $\bul$ symbols. The second clock register is initialized the same way the clock register was initialized in the previous construction.
\begin{eqnarray}
	\ket{\psi_{0}} &=& \quad \left[ \begin{array}{cccccccccccccccc}
		\tur & \bul &
		\bul & \bul & \bul & \bul & \bul & \sga & \wga & \iga & \iga & 
		\wga & \sga & \iga & \leftarrow & \tur
		\\		
		0 & 1    & 0    & 0    & 0    & 1    & 
		w_1 & w_2 & w_3
		 &
		1    & 0    & 0    & 0    & 1	& 0	& 0
		\\
		\bul & 0 & 0 & 0 & 0 & 0 & 0 & 0 & 0 & 0 & 0 & 0 & 0 & 0 & 0 & 0
		\\
		\bul & \bul & \bul & \bul & \bul & \bul & \bul & \bul & \bul & \bul & \bul & \bul & \bul & \bul & \bul & X
		\\
		\bul & \multicolumn{15}{c}{binary~x}
		\\
		\bul & 0 & 1 & 1 & 1 & 1 & 1 & 1 & 1 & 1 & 1 & 1 & 1 & 1 & 1 & 1
 	\end{array} \right]. \label{psi03}
\end{eqnarray}
\subsection{Transition Rules}
Here are the added transition rules, including a modification of transition rule 21:
\begin{eqnarray}
	\begin{array}{rccccrccc}
	21\,:& 
		\eight{\tur}{\tur}{-}{-}{-}{-}{\bul}{CX}
		&\goes&
		\eight{\leftarrow}{\tur}{-}{-}{-}{-}{\bul}{X}
	\end{array}
\end{eqnarray}
\begin{eqnarray}
	\begin{array}{rccccrccc}
	22\,:& 
		\eight{\tur}{\tur}{-}{-}{-}{-}{\bul}{C^{\times}}
		&\goes&
		\eight{\leftarrow^{\times}}{\tur}{-}{-}{-}{-}{\bul}{X}
	\end{array}	\label{rule22}
\end{eqnarray}
\begin{eqnarray}
	\begin{array}{rccccrccc}
	23a\,:& 
		\ten{\tur}{\tur}{not~\bul}{a}{\bul}{C}{not~\bul}{a}
		&\goes&
		\ten{\tur}{\tur}{not~\bul}{a}{\overleftarrow{C}}{\bul}{not~\bul}{a}
	&\quad&
	23b\,:& 
			\ten{\tur}{\tur}{not~\bul}{a}{\bul}{C}{\bul}{a}
			&\goes&
			\ten{\tur}{\tur}{not~\bul}{a}{\overleftarrow{C}}{\bul}{\bul}{a}
	\end{array}
\end{eqnarray}
\begin{eqnarray}
	\begin{array}{rccccrccc}
	23c\,:& 
		\ten{\tur}{\tur}{\bul}{a}{\bul}{C}{not~\bul}{a}
		&\goes&
		\ten{\tur}{\tur}{\bul}{a}{\overleftarrow{C}}{\bul}{not~\bul}{a}
	\end{array}\nonumber
\end{eqnarray}
\begin{eqnarray}
	\begin{array}{rccccrccc}
	24a\,:& 
		\ten{not~\tur}{-}{not~\bul}{a}{\bul}{\overleftarrow{C}}{not~\bul}{a}
		&\goes&
		\ten{not~\tur}{-}{not~\bul}{a}{\overleftarrow{C}}{\bul}{not~\bul}{a}
	&\quad&
	24b\,:& 
		\ten{not~\tur}{-}{not~\bul}{a}{\bul}{\overleftarrow{C}}{\bul}{a}
		&\goes&
		\ten{not~\tur}{-}{not~\bul}{a}{\overleftarrow{C}}{\bul}{\bul}{a}
	\end{array}
\end{eqnarray}
\begin{eqnarray}
	\begin{array}{rccccrccc}
	24c\,:& 
		\ten{not~\tur}{-}{\bul}{a}{\bul}{\overleftarrow{C}}{not~\bul}{a}
		&\goes&
		\ten{not~\tur}{-}{\bul}{a}{\overleftarrow{C}}{\bul}{not~\bul}{a}
	\end{array}\nonumber
\end{eqnarray}
\begin{eqnarray}
	\begin{array}{rccccrccc}
	25\,:& 
		\ten{\tur}{-}{-}{a}{\bul}{C}{-}{\overline{a}}
		&\goes&
		\ten{\tur}{-}{-}{a}{\bul}{CX}{-}{\overline{a}}
	\end{array}
\end{eqnarray}
\begin{eqnarray}
	\begin{array}{rccccrccc}
	26\,:& 
		\ten{not~\tur}{-}{-}{a}{\bul}{\overleftarrow{C}}{-}{\overline{a}}
		&\goes&
		\ten{not~\tur}{-}{-}{a}{\bul}{CX}{-}{\overline{a}}
	\end{array}
\end{eqnarray}
\begin{eqnarray}
	\begin{array}{rccccrccc}
	27\,:& 
		\ten{-}{-}{-}{a}{CX}{\bul}{-}{a}
		&\goes&
		\ten{-}{-}{-}{a}{\bul}{CX}{-}{a}
	\end{array}
\end{eqnarray}
\begin{eqnarray}
	\begin{array}{rccccrccc}
	28\,:& 
		\ten{-}{-}{\bul}{a}{\bul}{\overleftarrow{C}}{\bul}{a}
		&\goes&
		\ten{-}{-}{\bul}{a}{\overleftarrow{C}}{\bul}{\bul}{a}
	\end{array}
\end{eqnarray}
\begin{eqnarray}
	\begin{array}{rccccrccc}
	29\,:& 
		\ten{-}{not~\tur}{\bul}{\bul}{\bul}{\overleftarrow{C}}{\bul}{\bul}
		&\goes&
		\ten{-}{not~\tur}{\bul}{\bul}{\overleftarrow{C}}{\bul}{\bul}{\bul}
	\end{array}
\end{eqnarray}
\begin{eqnarray}
	\begin{array}{rccccrccc}
	30\,:& 
		\twelve{-}{\tur}{\bul}{\overleftarrow{C}}{0}{1}
		&\goes&
		\twelve{-}{\tur}{\bul}{R^{\times}}{0}{1}
	\end{array}
\end{eqnarray}
\begin{eqnarray}
	\begin{array}{rccc}
	\raiseonebox 
	31\,:& 
		\band{\rightarrow^{\times}}{\aga} 
		&\goes&
		\band{\bul}{\overrightarrow{A}^{\times}} 
		\\
	\end{array}
	\label{rule1}
\end{eqnarray}
\begin{eqnarray}
	\begin{array}{rccc}
	\raiseonebox 
	32\,:& 
		\band{\overrightarrow{A}^{\times}}{\bga} 
		&\goes&
		\band{\aga}{\rb^{\times}}
		\\
	\end{array}
	\label{rule2}
\end{eqnarray}
\begin{eqnarray}
	\begin{array}{rccc}
	\raiseonebox 
	33\,:& 
		\band{\overrightarrow{A}^{\times}}{\bul} 
		&\goes&
		\band{\aga}{\rightarrow^{\times}} 
		\\
	\end{array}
	\label{rule3}
\end{eqnarray}
\begin{eqnarray}
	\begin{array}{rccccrccc}
	34\,:&
		\band{\rightarrow^{\times}}{\bul} 
		&\goes&
		\band{\mov^{\times}}{\bul}
	\end{array} \label{rule4}
\end{eqnarray}
\begin{eqnarray}
	\begin{array}{rccccrccc}
	35\,:& 
		\band{\aga}{\mov^{\times}}
		&\goes&
		\band{\mov^{\times}}{\aga}
	\end{array}	\label{rule5}
\end{eqnarray}
\begin{eqnarray}
	\begin{array}{rccccrccc}
	36\,:& 
		\band{\bul}{\mov^{\times}}
		&\goes&
		\band{\bul}{\rightarrow^{\times}}
	\end{array}	\label{rule6}
\end{eqnarray}
\begin{eqnarray}
	\begin{array}{rccc}
	\raiseonebox 
	37\,:& 
		\band{\aga}{\leftarrow^{\times}} 
		&\goes&
		\band{\la^{\times}}{\bul} 
		\\
	\end{array}
	\label{rule7}
\end{eqnarray}
\begin{eqnarray}
	\begin{array}{rccc}
	\raiseonebox 
	38\,:& 
		\band{\bga}{\la^{\times}} 
		&\goes&
		\band{\lb^{\times}}{\aga}
		\\
	\end{array}
	\label{rule8}
\end{eqnarray}
\begin{eqnarray}
	\begin{array}{rccc}
	\raiseonebox 
	39\,:& 
		\band{\bul}{\la^{\times}} 
		&\goes&
		\band{\leftarrow^{\times}}{\aga} 
		\\
	\end{array}
	\label{rule9}
\end{eqnarray}
\begin{eqnarray}
	\begin{array}{rccc}
	\raiseonebox 
	40\,:& 
		\band{\bul}{\leftarrow^{\times}} 
		&\goes&
		\band{\bul}{\rmov^{\times}} 
		\\
	\end{array}
	\label{rule10}
\end{eqnarray}
\begin{eqnarray}
	\begin{array}{rccc}
	\raiseonebox 
	41\,:& 
		\band{\rmov^{\times}}{\aga} 
		&\goes&
		\band{\aga}{\rmov^{\times}} 
		\\
	\end{array}
	\label{rule11}
\end{eqnarray}
\begin{eqnarray}
	\begin{array}{rccc}
	\raiseonebox 
	42\,:& 
		\band{\rmov^{\times}}{\bul} 
		&\goes&
		\band{\leftarrow^{\times}}{\bul} 
		\\
	\end{array}
	\label{rule12}
\end{eqnarray}
\begin{eqnarray}
	\begin{array}{rccccrccc}
	43a\,:& 
		\band{\rightarrow^{\times}}{\tur}
		&\goes&
		\band{\tur}{\downdownarrows^{\times}}
	&\quad&
	43b\,:& 
		\band{\tur}{\leftarrow^{\times}}
		&\goes&
		\band{\tur}{\rightarrow^{\times}}
	\end{array}	\label{rule13}
\end{eqnarray}
\begin{eqnarray}
	\begin{array}{rccccrccc}
	44\,:& 
		\twelve{\tur}{\downdownarrows^{\times}}{\bul}{X}{-}{-}
		&\goes&
		\twelve{\tur}{\tur}{\bul}{L^{\times}}{-}{-}
	\end{array}	\label{rule14}
\end{eqnarray}
\begin{eqnarray}
	\begin{array}{rccccrccc}
	45\,:& 
		\twelve{\tur}{\tur}{\bul}{L^{\times}}{-}{0}
		&\goes&
		\twelve{\tur}{\tur}{\bul}{C^{\times}}{-}{1}
	\end{array}	\label{rule15}
\end{eqnarray}
\begin{eqnarray}
	\begin{array}{rccccrccc}
	46\,:& 
		\twelve{\tur}{\tur}{\bul}{L^{\times}}{0}{1}
		&\goes&
		\twelve{\tur}{\tur}{\bul}{C^{\times}}{1}{0}
	\end{array}	\label{rule16}
\end{eqnarray}
\begin{eqnarray}
	\begin{array}{rccccrccc}
	47\,:& 
		\twelve{-}{-}{\bul}{L^{\times}}{1}{1}
		&\goes&
		\twelve{-}{-}{L^{\times}}{\bul}{1}{1}
	\end{array}	\label{rule17}
\end{eqnarray}
\begin{eqnarray}
	\begin{array}{rccccrccc}
	48\,:& 
		\twelve{not~\tur}{-}{\bul}{L^{\times}}{0}{1}
		&\goes&
		\twelve{not~\tur}{-}{\bul}{R^{\times}}{1}{0}
	\end{array}	\label{rule18}
\end{eqnarray}
\begin{eqnarray}
	\begin{array}{rccccrccc}
	49\,:& 
		\twelve{not~\tur}{-}{R^{\times}}{\bul}{0}{1}
		&\goes&
		\twelve{not~\tur}{-}{\bul}{R^{\times}}{0}{0}
	\end{array}	\label{rule21}
\end{eqnarray}
\begin{eqnarray}
	\begin{array}{rccccrccc}
	50\,:& 
		\twelve{\tur}{\tur}{R^{\times}}{\bul}{0}{1}
		&\goes&
		\twelve{\tur}{\tur}{\bul}{C^{\times}}{0}{0}
	\end{array}	\label{rule22}
\end{eqnarray}
Whenever just $a$ appears in a transition rule it must be one of $\{0,1\}$. Whenever $a$ and $\overline{a}$ appear, one of $a$ or $\overline{a}$ must be in $\{0,1\}$ and the remaining symbol is either the logical negation, or the $\bul$ symbol.
\subsection{Discussion}
Here's a brief outline of how this modified construction with these new transition rules works:
\begin{itemize}
	\item Transition rules 22-30 implement a new part of the clock update phase by giving it a new function at the end: when the clock pointer has finished updating the clock and is in $C$ mode, it then begins a second sweep to the left in $\overleftarrow{C}$ mode, comparing the number stored in the clock to the number stored in the target register bit by bit
	\item If the clock pointer discovers bits that do not match, it turns into $CX$ mode, moves back to the right, and the active symbol moves back up to the program layer to begin the next unitary application as normal in construction III
	\item If the clock pointer in $\overleftarrow{C}$ mode determines that the clock and target registers match, it continues moving left until it reaches the left end of the chain, and this causes the clock pointer to transition to $R^{\times}$ mode
	\item Once the the clock pointer has transitioned to $R^{\times}$ mode, all future active symbols will have the $\times$ symbol attached and obey transition rules 31-50, which identically mimic transition rules 1-20 of construction $III$, but are modified slightly so that a unitary is never applied to the work qubits ever again (the presence of the $\times$ symbol on all active symbols indicates this), and now the clock pointer (operating in an identical fashion to how it did in construction $III$, without a phase where it compares to a target register) treats the the second clock register $C2$ as the clock register.
\end{itemize}
At this point, based on our in depth analysis of construction III, it is easy to check that the sequence of states generated by forward transition rules $1-50$ is UOG by the appropriate sequence of unique forward transitions. The outcome, then, is that the Hamiltonian $H_{IV}$ based on transition rules 1-50 implements a quantum walk on a line whose length is $l={\rm{exp}}(N,K)$ (the length of the line for constructions $H_{III}$ and $H_{IV}$ is exponential because the clock can count up to a time ${\rm{exp}}(N,K)$). Once the quantum walk has proceeded far enough for $U$ to have been applied to the work qubits $x$ times (this distance is ${\rm{poly}}(x,N,K)$), the quantum walk can continue forward for ${\rm{exp}}(N,K)$ more steps without ever changing the state of the data and clock registers ever again. We complete the analysis of the construction below by proving that we only need to run the simulation for a time ${\rm{poly}}(x,N,K)$ in order to be able to do a simple measurement on the chain that will collapse the state of the work qubits to one where $U^x$ has been applied with high probability.
\subsection{Run-Time Analysis}\label{runtime}
Here we prove that after running construction $IV$ with an appropriate input state for a time $\tau={\rm{poly}}(x,N,K)$, a computational basis measurement on the clock register will return the result that $x$ is stored in the clock register (and therefore $U^x$ has been applied to the work qubits) with high probability.

This is accomplished by slightly modifying a lemma from \cite{NW}:
\begin{lem7}
	Consider a continuous-time quantum walk on a line of length $l$, where the Hamiltonian is the negative of the adjacency matrix for the line. Let the system evolve for a time $\tau\leq\tau^*$ chosen uniformly at random, starting from position basis state $\ket{0}$ (one end of the line). The probability to measure a state $\ket{m}$ with $m>(1-F)l$, $F\in[0,1]$, is then bounded from below as $p^*\geq F - O\left(\frac{l}{\tau^*}\right)-O\left(\frac{1}{l}\right)$.
\end{lem7}
\begin{proof}
Let $p_{\tau}(m)$ be the probability of measuring state $\ket{m}$ at time $\tau$ in the above scenario. Then the time average of $p_{\tau}(m)$ for $0\leq\tau\leq\tau^*$ is
\begin{eqnarray}
	\overline{p}_{\tau^*}(m)=\frac{1}{\tau^*}\int_{0}^{\tau^*}p_{\tau}(m)d\tau
\end{eqnarray}
which, in the limit $\tau^*\to\infty$, converges to the limiting distribution \cite{NW}
\begin{eqnarray}
	\pi(m)=\frac{2+\delta_{m,0}+\delta_{m,l-1}}{2(l+1)},
\end{eqnarray}
in the sense that
\begin{eqnarray}
	\sum_{m=0}^{l-1}|\overline{p}_{\tau^*}(m)-\pi(m)|\leq O\left(\frac{l}{\tau^*}\right).
\end{eqnarray}
Then, the probability of measuring state $\ket{m}$ for $m>(1-F)l$ for time $\tau<\tau^*$ chosen uniformly at random is
\begin{eqnarray}
	p^* = \sum_{m>(1-F)l} \overline{p}_{\tau^*}(m).
\end{eqnarray}
We have that
\begin{eqnarray}
	\sum_{m=0}^{l-1}|\overline{p}_{\tau^*}(m)-\pi(m)|&\geq& \sum_{m>(1-F)l}|\overline{p}_{\tau^*}(m)-\pi(m)|
	\\
	&\geq&\left| \sum_{m>(1-F)l}\left(\overline{p}_{\tau^*}(m)-\pi(m)\right) \right|
	\\
	&=&\left| p^* - \sum_{m>(1-F)l} \frac{2+\delta_{m,0}+\delta_{m,l-1}}{2(l+1)} \right|
	\\
	&=&\left| p^* - F+\frac{F}{l+1}+\frac{1}{2(l+1)}\right|
	\\
	&\geq&-p^* + F-\frac{F}{l+1}-\frac{1}{2(l+1)}
\end{eqnarray}
But, $O\left(\frac{l}{\tau^*}\right)\geq\sum_{m=0}^{l-1}|\overline{p}_{\tau^*}(m)-\pi(m)|$, thus, we have that 
\begin{eqnarray}
	p^*\geq F - O\left(\frac{l}{\tau^*}\right)-O\left(\frac{1}{l}\right),
\end{eqnarray}
as desired.
\end{proof}
We now consider the consequences of this lemma for the run-time of our simulation. The length of the chain that the states $\ket{\psi_t}$ of the simulator map onto position eigenstates $\ket{t}$ of is $l={{\rm{exp}}}(N,K)$. For $x={\rm{exp}}(N)$, parameters can easily be chosen such that the fraction $F$ of $l$ such that $\ket{\psi_{t> (1-F)l}}$ has had $U^x$ applied to the work qubits is $1-{\rm{exp}}(-N)$. Then by setting $\tau^*={\rm{poly}}(l)$ appropriately, and picking a uniformly random time between $0$ and $\tau^*$, we will succeed in applying $U^x$ to the work qubits with probability $p^*\geq 1-1/{\rm{exp}}({\rm{poly}}(N,K))$. For all states $\ket{\psi_{t> (1-F)l}}$, which have had $U^x$ applied to the work qubits, the clock register will hold the number $x$ which was originally set in the target register. So we only need to do a computational basis measurement on the clock register and find it in the state holding the binary number $x$ to know that we've succeeded.

Note that if we would like to simulate the application of unitary $U^x$ for $x={\rm{poly}}(N)$, the above strategy does not actually seem very helpful: because $l={\rm{exp}}({\rm{poly}}(N,K))$ is exponential, we need to wait for, on average, exponential time to guarantee high success probability even though we're simulating something that should only take polynomial time. This can easily be fixed. By modifying the start state of the simulator, $\ket{\psi_0}$, so that the $\bul$ symbol in the target register is far enough to the right, one can exponentially reduce $l$. This is because once the target register starts functioning as a clock (after $U^x$ has been applied), the length of the remaining distance the particle can travel is exponential in the number of sites to the right of the $\bul$ symbol in the target register. The end state is not reached until the target register has counted up to the largest number that it can.  So, one can place the $\bul$ at the appropriate site (say, 3 sites to the left of the most significant digit of the number $x$ stored in the target register) so that $l$ will be polynomial in $N$, but will also be large enough relative to $x$ so that one only needs $\tau^*$ to be polynomial to achieve a high rate of success. Note, however, that in this scenario, the probability of success will only be inverse polynomially close to unity rather than exponentially.

Finally, we mention that from the start state defined in equation (76), it is actually possible to apply reverse transitions to this start state which allows you to start running a `clock check and update' phase in reverse using the reverse of transition rules 21, 25 and 20 (if the target register's least significant bit is 1), or rules 21, 27, 23, 24 and 20 (if the target register's least significant bit is 0). However, it is easy to check that in either case, only one unique sequence of no more than $3L$ reverse transitions can be applied to this state before reaching a state from which no further reverse transitions can be made. Thus the effect on run-time analysis is insignificant.

\section{Acknowledgements}

We thank Dorit Aharonov for interesting discussions, Elizabeth Crosson for helpful discussions and suggestions, and Toby Cubitt for helpful comments on our first draft that lead us to correct several mistakes in our construction. Author T. B. acknowledges financial support from the National Science and Engineering Research Council of Canada (NSERC) in the form of a Postgraduate Scholarship (PGS-D) award during the time in which this work was completed.

\bibliographystyle{unsrt}

\bibliography{hambib}

\end{document}